%% file: SoundP2Plncs.tex
\documentclass{llncs}
\usepackage[german,english]{babel}
\usepackage{times,theorem}
\usepackage{latexsym,color,graphics}
\usepackage{diagrams}
\usepackage{url}
\usepackage{epsfig}
\usepackage{amssymb}
\usepackage{amsmath}
\usepackage{amsfonts}

\input{macros-general1}
\input{macros-l1}
\title{Sound and Complete Query Answering in Intensional P2P Data Integration}
\author{Zoran Majki\'c}
\authorrunning{Zoran Majki\'c}
\institute{International Society for Research in Science and Technology \\
PO Box 2464 Tallahassee, FL 32316 - 2464 USA\\
\email{majk.1234@yahoo.com}\\ http://zoranmajkic.webs.com/}
\authorrunning{Zoran Majki\'c}
\newtheorem{theo}{Theorem}
\newtheorem{propo}{Proposition}
\newcount\pdfoutput
\begin{document}

% \firstpage{1}

\maketitle
\begin{abstract}
Contemporary use of the term 'intension' derives from the
traditional logical doctrine that an idea has both an extension and
an intension. In this paper we introduce an intensional FOL
(First-order-logic) for P2P systems by fusing the Bealer's
intensional algebraic FOL  with the S5 possible-world semantics of
the Montague, we
define the \emph{intensional equivalence} relation for this logic and the weak deductive inference for it.\\
The notion of ontology has become widespread in semantic Web. The
meaning of concepts and views defined over some database ontology
can be considered as intensional objects which have particular
extension in some possible world: for instance in the \emph{actual}
world. Thus, non invasive mapping between completely independent
peer databases in a P2P systems can be naturally specified by the
set of couples of intensionally equivalent views, which have the
same meaning (intension), over two different peers. Such a kind of
mapping has very different semantics from the standard view-based
mappings based on the material implication commonly used for Data
Integration. We show how a P2P database system may be embedded into
this intensional modal FOL, and how we are able to obtain a weak
non-omniscient inference, which can be effectively implemented.
 For a query answering we consider non omniscient
query agents and we define object-oriented class for them which
implements as method the query rewriting algorithm. Finally, we show
that this query answering algorithm is sound and complete w.r.t. the
weak deduction of the P2P intensional logic.
\end{abstract}

%\begin{keywords}
%Keywords: Inovative Web-based Applications, Semantic Web,Web Query
%Language,Web Information retrieval.
%\end{keywords}

%\newpage
%----------------------------------------------------------------------
% SECTION I: Introduction
%----------------------------------------------------------------------
\section{Introduction}

  Ontologies play a prominent role on the Semantic Web.  An ontology specifies a conceptualization of a domain
in terms of concepts, attributes and relations.
 However,
because of the Semantic Web distributed nature, data on it will
inevitably come from many different ontologies. A key challenge in
building the Semantic Web, one that has received relatively little,
attention, is finding semantic mappings among the ontologies
(peers).  Given the de-centralized nature of the development of the
Semantic Web, there will be an explosion in the number of
ontologies. Many of these ontologies will describe similar domains,
but using different terminologies, and others will have overlapping
domains. To integrate data from disparate ontologies, we must know
the \emph{semantic correspondence} between their elements
~\cite{Usho01}. Recently are given a number of different
architecture solutions
~\cite{GHIR01,SGMB01,GhGi01,Reit84,Majk03s,CDGL04}. The authors provided more information about different P2P systems
and a comparative analysis in \cite{Majk06J,MaPr08}. \\
 In what follows we will consider
a reach ontology of peer
  databases, formally expressed as a global schema of a Data
  Integration System (DIS).
  A DIS ~\cite{Lenz02} is a triple $\I_i
= (\G_i,\S_i,\M_i)$, where $\G_i = (\O_i, \Sigma_{T_i})$ is a global
schema, expressed in a language $\L_{\O}$ over an alphabet
$\A_{\G_i}$, $~\Sigma_{T_i} $ are the integrity constraints, $\S_i$
is a source schema and $\M_i$ is a set of mappings between a global
relational database schema (ontology) $\O_i$ and a source relational
schema $\S_i$ of data extracted by wrappers. In what follows we will
consider the case of Global-As-View (GAV) mappings between source
and global schema, with existential quantifiers also (for mapping an
incomplete information from source to global schema). A DIS with
constraints (for example, key-constraints) can become \emph{locally
inconsistent} w.r.t. data sources extracted by wrappers. Such local
inconsistences can be avoided by inconsistency
repairing technics \cite{GrGZ01}.  \\
We conceive a peer $P_i$ as a software module, which encapsulates a
DIS $~\I_i$. The internal structure of a peer database is hidden to
the user, encapsulated in the way that only its logical relational
schema $\O_i$  can be seen by users, and is able to respond to
the union of conjunctive queries by \emph{known} answers (true in all models of a peer-database).\\
We consider \emph{a view} definition $q_k(\textbf{x})$ as a
conjunctive query, with a tuple of variables in $\textbf{x}$,
$~~head (q_k) \leftarrow body(q_k)~~$ where $~body(q_k)~$ is a
sequence $b_1, b_2,..., b_m$, where each $b_j$ is an atom over a
global relation name of a peer $P_i$. In what follows we will
consider a view as a virtual predicate with a tuple $\textbf{x}$ of free variables in the head of a query.\\
% Interested reader can see \cite{Majk05p},
%where was shown that \emph{plausible} query answering, such that is
%true in each preferred model of a peer database, corresponds to the
%cumulative non-monotonic inference; in the limit case when we
%consider \emph{all} minimal Herbrand models, we will have the case
%of \emph{known} query answering.\\
 In P2P systems, every node (peer)
of the system acts as both client and server and provides part of
the overall information available from an Internet-scale distributed
environment. In this paper we consider a formal framework based on
the following considerations ~\cite{MaLe03}: what we need here is\\
%\begin{itemize}
 -  a mechanism that is able, given any
  two peer databases, to define mappings between them, without
  resorting to any unifying (global) conceptual structure.\\
  -  a completely decentralized network of database
  peers: all peers serve as entry points for search, offering
  their relational schema in order to formalize user queries.\\
  - query answering, fundamentally based on interactions
  which are strictly local and guided by locally defined mappings
  of a considered peer w.r.t. other peers.\\
  - not limit a-priori the topology of the
  mapping assertions between peers in the system: we do not impose acyclicity of assertions.\\
  - a semantic characterization that leads to
  setting where query answering is decidable, and possibly,
  polynomially tractable.\\
  %: that is a reason to use epistemic modal  instead of FOL  to model.
%\end{itemize}
The last two considerations, decidability and non acyclicity enforce
the reason to use \emph{epistemic modal} instead of FOL
(First-Order-Language; see for example \cite{CDGL04}) to model a
peer database, with an epistemic operator "peer $P_i$ knows",
$~K_i$, and with relational database schema (ontology)
 $\O_i$ for \emph{conjunctive} query language.
Such an epistemic semantics of peers has been  presented originally
in \cite{Majk03s} based on the \emph{hybrid mono-modal} language
 with
a unique universal modal operator $\square$ ~\cite{Blac00}, so that
where $~K_i = @_i\square~$, where new modal operator, $@$, for this
hybrid logic enables to "retrieve" worlds: A formula of the form
$@_i \varphi$ is an instruction to \emph{move} to the world labeled
by the variable $~i~$ and evaluate $\varphi$ there.
 How this
relational, view based, approach to peer ontologies relates to the
actual semantic Web RDF language is explained in
\cite{DeHS04,Majk05w,Majk06fl,Majk08rdf}.\\
 In what follows we abbreviate $~~A\Rightarrow
B \wedge B\Rightarrow A~~$ by $~~A \equiv B$.\\
 Let
$q_{P_i}(\textbf{x})$ and $q_{P_k}(\textbf{x})$ be  two views
(conjunctive queries) over $P_i$ and $P_k$ respectively, with the
same set of free variables
$\textbf{x}$, then we can have two P2P scenarios:\\
1. The \emph{strong (extensional)} multi-modal mapping, introduced
%in \cite{Majk03s},
 by a  formula
$~~~K_i q_{P_i}(\textbf{x})\Rightarrow K_k q_{P_k}(\textbf{x})~~$,
where $'\Rightarrow'$ is the logic implication, used in a single S5
modality ~\cite{FKLS03r,CDGL04,CGLMR05}, and  in K45 multi-modality
\cite{CGLMR05}. It tells that the knowledge of the peer $P_i$
contained in its view $q_{P_i}(\textbf{x})$ \emph{must be} contained
in the view $q_{P_k}(\textbf{x})$ of the peer $P_k$. In this case
this forced transfer of the local data of one peer to other peers
can render inconsistent knowledge of other peers, and from the
semantic point of view, reassembles the kind of strong data
integration system, with a global logic of the whole P2P system and
a recursive Datalog for query rewriting \cite{CDGL04}.\\
%This is very good solution for classical data integration where somebody knows
%all about all, but not for the chaotic Web P2P system, where each
%peer can change in any moment also its proper ontology.\\
2. The \emph{weak (intensional)} mapping, defined by the "formula"
$~K_i q_{P_i}(\textbf{x}) \approx_{in} K_k q_{P_k}(\textbf{x})$,
$~~$ where $'\approx_{in}'$ is the informal symbol for the
intensional equivalence
~\cite{Majk06J,Majk03s,Majk03q,Majk03p,Majk04ph}, and formally  in
Definition \ref{def:equival} by the logic modal formula
$~~\diamondsuit q_{P_i}(\textbf{x}) \equiv \diamondsuit
q_{P_k}(\textbf{x})$ of the intensional FOL introduced in this
paper. This mapping tells only that these two peers have the
knowledge about the \emph{same} concept, without any constraint for
 extensions of this concept
in these two peers respectively. \\
%In what follows we will consider only this intensional version for
%P2P mapping based on considerations presented in \cite{Majk04ph}.
The more complete comparative analysis of these two different
approaches can be found in \cite{Majk06J}. In what follows we will
consider only this \emph{new intensional version} for P2P mapping
better suited for fully independent peers \cite{Majk04ph}.
%differently form the Global-Local-as-view (GLAV) strong mappings,
%here the deductive system of the modal S5 intensional FOL, given a
%user query, derives all intensionally equivalent queries over other
%peers,and there is \emph{no any transfer} of data from one peer to other, which can introduce inconsistent information to other peers.\\
     Consequently, in order to be able to share the knowledge with
other peer $P_j$ in the network $\N$, each peer $P_i$ has also an
export-interface module $\M^{ij}$ composed by groups of ordered
pairs of intensionally equivalent  views (conjunctive queries over
peer's ontologies), denoted by $(q_i,q_j)$.
\begin{definition} \label{def:P2Pnetwork} \cite{Majk03s}
The P2P network system $\N$ is composed by $2\leq N$ independent
peers, where each peer module $P_i$
%, based on the ontology $\O_i$ (an epistemic S5 normal modal logic, where  possible worlds are
% minimal Herbrand models of a peer database with incomplete
%information, with epistemic operator "know" $K_i$, and with
%relational database schema interface $\O_i$ for conjunctive query language),
is defined as follows:
%\\ \begin{center}
$~~~~~~P_i := \langle \O_i, ~\M_i \rangle $, $~~~~~~$ where \\$\M_i
= <\M^{i1},...,\M^{iN}>$ is an interface tuple with
%\end{center}
  $\M^{ij}, ~1\leq j\leq N$ a (possibly empty) interface to other peer
$P_j$ in the network, defined as a group of intensionally equivalent
query connections, denoted by $(q^{ij}_{1k}, q^{ij}_{2k})$ where
$q^{ij}_{1k}(\textbf{x})$ is a conjunctive query defined over
$\O_i$, while $q^{ij}_{2k}(\textbf{x})$ is a conjunctive query
defined over the ontology $\O_j$ of the connected peer $P_j~$ :
%\begin{center}
$~~~\M^{ij} = \{(q^{ij}_{1k}, q^{ij}_{2k})~|~ ~~1\leq k \leq n_{ij} \}$,\\
%\end{center}
where $n_{ij}$ is the total number of query connections of the peer
$P_i$ toward a peer $P_j$.
\end{definition}
Intuitively, when an user defines a conjunctive query over the
ontology $\O_i$ of the peer $P_i$, the intensionally equivalent
concepts between this peer and other peers will be used in order to
obtain the answers from a P2P system. \\They will be the "bridge"
which a query agent can use to rewrite the original user query over
a peer $P_i$ into \emph{intensionally equivalent} query over other
peer $P_j$ which has different (and independent) ontology from the
peer $P_i$.\\ The answers of other peers will be epistemically
considered as \emph{possible} answers because the are based on the
\emph{belief} which has the peer $P_i$ about the knowledge of a peer
$P_j$: this belief is formally represented by supposition of a peer
$P_i$ that the pair of queries $(q^{ij}_{1k}, q^{ij}_{2k}) \in
\M^{ij}$ is
intensionally equivalent.\\
\textbf{Motivation:} The main motivation for the introduction of
intensional logic for the mappings between peers is based on the
desire to have the \emph{full} epistemic independency of peer
databases: we consider that they can change their ontology and/or
extension of their knowledge independently from other peers and
without any communication to other peers. So, we intend to use the
mappings between peers that are not controlled by any centralized
system, which are not permanently correct during evolution of a P2P
system in time but express only assumptions based on their local
belief about knowledge of other peers \cite{Majk06J}. Here there is
no any transfer of the local knowledge of a peer to knowledge of
other peers, which can possibly generate inconsistency of these
other peers, but only a belief based assumption that they can speak
about intensionally equivalent concepts. From a practical point of
view, we assume that there is no any \emph{omniscient} query agent,
able to know the whole global P2P system. Consequently, as in human
communications, based on the fact that the same concepts have the
same meaning for people, but not the same extensions for every human
being, a query answering must be based on the weaker form of
deduction that the omniscient deduction which uses Modus Ponens and
Necessity rule (for normal modal logic) to derive all possible
deductions. \\ The formalization of this non omniscient intensional
contextual reasoning for the query-agents in P2P database systems is
presented in \cite{Majk06Om}. In this way we intend to obtain the
very robust P2P systems but also the possibility to map naturally
P2P database systems into grid computations: if the peers are fully
independent it is enough to associate each pair (peer, query
formulae) to a particular resource of grid computing,
in order to obtain known answer from such a peer.\\
The aim of this paper is to provide the clear \emph{semantics} for
such P2P database systems with intensional mappings between peers,
%based on the fact that each peer is a database with incomplete and
%partially inconsistent information (thus with a number of minimal
%Herbrand models), but able to provide known answers based on its
%epistemic S5 modal logic,
and to provide the clear \emph{mathematical} framework for its query
answering computation which, successively, can be implemented into
an massive grid
computing framework.\\
The main contributions in this paper are the following:\\
1. We define a modal logic framework:  we define an intensional S5
modal FOL  $\L_{\omega}$ with intensional identity for a P2P system,
by fusing Bealer's algebraic and Montague's possible world
approaches, and enrich it with the \emph{intensional equivalence}.
We define a \emph{weak deduction} inference for this intensional
logic to be implemented by query answering algorithms of non
omniscient query agents. This logic is S5 modal logic where the set
of possible worlds  is the set of all possible evolutions in a time
of a given P2P system (when the peers
modify their ontologies or their  extensions). \\
2. Finally, we define an object-oriented class for query agents
which implements, as method, a query rewriting algorithm able to
reformulate the original user conjunctive query specified over a
peer $P_i$, in intensionally equivalent queries for  other peers. We
show that this algorithm is sound and complete w.r.t. the weak
deduction of the intensional logic $\L_{\omega}$.
\\
This paper is written to be selfcontained, so the original part
w.r.t the previous publications cited by author  is presented in the
last Section 5.
\\
The Plan of this work is the following: in Section 2 is presented
  the formal semantics for intensional FOL  and intensional equivalence,
   used to define a non invasive semantic mappings between epistemically independent peers, based on relational views of different
   peers. This Section is a fundamental background for the rest of the
   paper, and distinguish this approach from all other currently
   used for definition of mappings between peers, as remarked in the
   introduction. It combines the Bealer's algebraic  and Montague's
   possible worlds semantics for intensional logic FOL: The Bealer's algebra is useful in order to define the abstraction of logic formulae,
   in order to be used as terms
   in other logic formulae, while the Montague's possible worlds semantics is used to  define the
   intensional   equivalence  by existentially quantified modal
   formulae.\\
     In Section 3 we define an embedding of P2P systems into this intensional FOL with standard S5 modal omniscient inference.
     In Section 4 we define the weak (non-omniscient) deduction inference for it, different from the omniscient inference
     of the intensional FOL, which can be effectively used by query
     agents for computation considering only the actual Montague's
     world.
  Finally, in
  Section 5 we define the weak deduction solution for query answering in this intensional logic, and
  we define the non omniscient query agent object-oriented class which implements sound and complete query rewriting algorithm w.r.t.
  the weak intensional deduction.
  %$\vspace*{-4mm}$
%
%\section{ P2P network definition: Syntax \cite{Majk03s}}}
%

%  \textbf{P2P network definition: Syntax \cite{Majk03s}}\\

% The query rewriting in a cyclic P2P system will be described in the
% Example 2 and Fig.2.
%
%$\vspace*{-3mm}$
%\subsection{Introduction to Intensional First-order Logic}
\section{Intensional equivalence and Intensional FOL language} \label{section:intensional}
%\label{section:intensional}
%
Contemporary use of the term 'intension' derives from the
traditional logical doctrine that an idea has both an extension and
an intension. Intensional entities are such things as concepts,
propositions and properties. What make them 'intensional' is that
they violate the principle of extensionality; the principle that
extensional equivalence implies identity. All (or most) of these
intensional entities have been classified at one time or another as
kinds of
Universals \cite{Beal93}.\\
 The fundamental entities
are \emph{intensional abstracts} or so called, 'that'-clauses. We
assume that they are singular terms; Intensional expressions like
'believe', mean', 'assert', 'know',
 are standard two-place predicates  that take 'that'-clauses as
 arguments. Expressions like 'is necessary', 'is true', and 'is
 possible' are one-place predicates that take 'that'-clauses as
 arguments. For example, in the intensional sentence "it is
 necessary that A", where $A$ is a proposition, the 'that A' is
 denoted by the $\langle A \rangle$, where $\langle \gtrdot$ is the intensional abstraction
 operator which transforms a logic formula into a \emph{term}. So
 that the sentence "it is
 necessary that A" is expressed by the logic atom $N(\langle A \rangle)$, where
 $N$ is the unary predicate 'is necessary'. In this way we are able
 to avoid to have the higher-order syntax for our \emph{intensional} logic language
 (predicates appear in variable places of other predicates),as, for example HiLog \cite{ChKW93} where the \emph{same}
 symbol may denote a predicate, a function, or an atomic formula. In
 the First-order logic (FOL) with intensional abstraction we have
 more fine distinction between an atom $A$ and its use as a
 \emph{term} "that A", denoted by $\langle A \rangle$ and considered
 as intensional "name",  inside some other predicate, and, for example, to have the first-order formula $\neg A \wedge P(t, \langle A
 \rangle)$ instead of the second-order HiLog formula $\neg A \wedge P(t,  A
 )$ .
 %$\vspace*{-3mm}$
 \begin{definition} \label{def:bealer}
  The syntax of the First-order Logic language with intensional abstraction
$\langle \rangle$, called $\L_{\omega}$ in \cite{Beal79}, is as follows:\\
 Logic operators $(\wedge, \neg, \exists)$; Predicate letters in $P$
 (functional letters are considered as particular case of predicate
 letters); Variables $x,y,z,..$ in $Var$; Abstraction $\langle \_ \rangle$, and punctuation
 symbols (comma, parenthesis).
 With the following simultaneous inductive definition of \emph{term} and
 \emph{formula}   :\\
 %\begin{enumerate}
   1. All variables and constants (0-ary functional letters in P) are terms.\\
   2. If $~t_1,...,t_k$ are terms, then $A(t_1,...,t_k)$ is a formula
 ($A \in P$ is a k-ary predicate letter).\\
   3. If $A$ and $B$ are formulae, then $(A \wedge B)$, $\neg A$, and
 $(\exists x)A$ are formulae.\\
   4. If $A$ is a formula and $\alpha = <x_1,...,x_n>$, is a sequence (tuple) of \emph{distinct} variables (a subset of free variables in A),
 then $\langle A \rangle_{\alpha}$ is a term. The externally quantifiable variables are the \emph{free} variables not in $\alpha$.
  When $n =0,~ \langle A \rangle$ is a term which denotes a
proposition, for $n \geq 1$ it denotes
 a n-ary relation-in-intension.\\
% \end{enumerate}
An occurrence of a variable $x_i$ in a formula (or a term) is
\emph{bound} (\emph{free}) iff it lies (does not lie) within a
formula of the form $(\exists x_i)A$ (or a term of the form $\langle
A \rangle_{x_1...x_i...x_m}$). A variable is free (bound) in a
formula iff it has (does not have) a free occurrence in that
formula.\\ A \emph{sentence} is a formula having no free variables.
The binary predicate letter $F_1^2$ is singled out as a
distinguished logical predicate and formulae of the form
$F^2_1(t_1,t_2)$ are to be rewritten in the form $t_1 = t_2$. The
logic operators $\forall, \vee, \Rightarrow$ are defined in terms of
$(\wedge, \neg, \exists)$ in the usual way.
%$\vspace*{-3mm}$
\end{definition}
For example, "x believes that A" is given by formula $B(x,\langle A
\rangle)$ ( $B$ is binary 'believe' predicate), "Being a bachelor is
the same thing as being an unmarried man" is given by identity of
terms $\langle B(x) \rangle_x = \langle U(x) \wedge M(x) \rangle_x$
(with $B$ for 'bachelor', $U$ for 'unmarried', and $M$
for 'man', unary predicates).\\
%$\vspace*{-3mm}$
%
%\section{Intensional equivalence and Intensional FOL language} \label{section:intensional}
%
% We will extend the intensional FOL language of Bealer, given in the introduction, by Definition \ref{def:bealer}:
 %,  by other three operators for intensional entities:\\
%1. Algebraic conjunction:$~~\langle A \rangle conj \langle B
%\rangle = \langle A \wedge B \rangle$;\\
%2. Algebraic disjunction:$~~\langle A \rangle disj \langle B
%\rangle = \langle A \vee B \rangle$;\\
%3. Algebraic implication:$~~\langle A \rangle impl \langle B
%\rangle = \langle A \Rightarrow B \rangle$;\\
%4. Algebraic negation:$~~neg \langle A \rangle  = \langle \neg A
% \rangle$.\\
Thus, analogously to Boolean algebras
 which are extensional models of propositional logic, we introduce an
 intensional algebra as follows. We consider a non empty domain $~\D = D_{-1} \bigcup D_I$,  where a subdomain $D_{-1}$ is made of
 particulars (extensional entities), and the rest $D_I = D_0 \bigcup
 D_1 ...\bigcup D_n ...$ is made of
 universals ($D_0$ for propositions (the 0-ary relation-in- intensions), and  $D_n, n \geq 1,$ for
 n-ary relations-in-intension.
 % $\vspace*{-3mm}$
\begin{definition}  \label{def:Syntax} Intensional algebra is a structure \\$~Alg_{int} =
~~<\D, conj, disj, impl, neg, pred,\tau, f,t
 >$, $~~~~~~$ with
%which contains  a non empty domain $~\D $;\\
 binary operations\\  $conj:D_I\times D_I \rightarrow D_I$,  $~~~ pred:D_i\times \D \rightarrow D_{i-1}$, $~~$ for $i \geq 1$,
  $~~$ and unary operation  \\$~~neg:D_i\rightarrow D_i$,  for each $i \geq
0$; the disjunctions and implications are defined in a standard way
by $~~disj(u,v) = neg(conj(neg(u), neg(v)))$, $impl(u,v) =
disj(neg(u),v)$, for any $u,v \in D_I$;
 \\
 $\tau$  is a set of auxiliary operations \cite{Beal82} intended to be semantic
 counterparts of the syntactical operations of repeating the same
 variable one or more times within a given formula and of changing
 around the order  of the variables within a given formula;\\
 $f, t$
 % \in D_{-1}$
  are empty set $\{\}$ and set $\{<>\}$ (with the empty tuple $<> \in D_{-1}$ i.e. the unique tuple of 0-ary relation)
 which may be thought of
 as falsity and truth, as those used  in the relational algebra, respectively.
 %$\vspace*{-2mm}$
\end{definition}
 \textbf{Remark}: This definition differs from  the original work in \cite{Beal82} where $t$ is defined as $\D$, and  $conj:D_i\times D_i \rightarrow
D_i$, $i \geq 0$, here we are using the \emph{relational algebra}
semantics for the conjunction. So that we are able to support also
structural composition for abstracted terms  necessary for
supporting relational conjunctive queries, as, for example, $\langle
A(x,y) \wedge B(y,z) \rangle_{xyz}$, which is not possible in the
reduced syntactic version of the Bealer's algebra. In the original
work \cite{Beal82} this "algebraization" of
 the intensional FOL is extended also to logic quantifiers,
 but for our purpose it is not necessary, because in the embedding of a P2P system into the intensional FOL for the query answering, we will use
 only the predicates from the global schema of each peer databases both with the queries (virtual predicates) used for intensional mapping
 between peers. The rest of peer's ontology (a Data Integration System) can use also existential quantifiers for internal mappings
  between source and global schema of a peer database (in the case of \emph{incomplete} information which comes from some source into relations of a
  global schema (in GAV mappings), or in the case of particular integrity constraints over a global schema of a peer
database).
   But as we noted in the introduction,
  this part of a peer ontology is encapsulated into the peers and is responsible only to define the exact \emph{extension},
  in a given instance of time (possible world
  for the intensional S5 modal FOL), for predicates used in the intensional FOL. In order to compute this extension, \emph{independently} for each peer
  database, we will use the ordinary \emph{extensional} FOL logic for encapsulated peers, based on the extensional S5 epistemic FOL (Subsection
  3.1). The mapping $V$, used in the following Montague's based
  approach, is a high-level result of the data semantics encapsulated
  into each peer database. It is logically specified in this
  extensional S5 epistemic FOL for a peer database.
  \\$\square$\\
%\vspace*{-4mm}
%
%\subsection{Semantics and Intensional equivalence \label{section:intensional}}
%
 The distinction between intensions and
extensions is important
 especially because we are now able to have and \emph{equational
 theory} over intensional entities (as  $\langle A \rangle$), that
 is predicate and function "names", that is separate from the
 extensional equality of relations and functions. Thus, intensional
 FOL  has the simple Tarski first-order semantics, with a decidable
 unification problem, but we need also the actual world mapping
 which maps any intensional entity to its \emph{actual world
 extension}. In what follows we will identify a \emph{possible world} by a
 particular mapping which assigns to intensional entities their
 extensions in such possible world. It is direct bridge between
 intensional FOL  and possible worlds representation
 \cite{Lewi86,Stal84,Mont70,Mont73,Mont74}, where the intension of a proposition is a
 \emph{function} from a set of possible worlds $\mathbb{W}$ to truth-values, and
 properties and functions from $\mathbb{W}$ to sets of possible (usually
 not-actual) objects.\\
 In what follows we will use one simplified S5 modal logic framework (we
will not consider the time as one independent parameter as in
Montague's original work) with a model $\M = (\mathbb{W}, \R,
\D,V)$, where $\mathbb{W}$  is a set of possible worlds, $\R$ is a
reflexive, symmetric and transitive accessibility relation between
worlds ($\R = \mathbb{W} \times \mathbb{W}$), $\D$ is a non-empty
domain of individuals given by Definition \ref{def:Syntax}, while
$V$ is a function defined
for the following two cases:\\
1. $~V:\mathbb{W} \times F \rightarrow \bigcup_{n < \omega}
\D^{\D^n}$, with $F$ a set of functional symbols of the language,
such that for any world $w \in  \mathbb{W}$ and a functional symbol
$f \in F$, we obtain
a function $V(w,f):\D^{arity(f)} \rightarrow \D$.\\
2. $~V:\mathbb{W} \times P \rightarrow \bigcup_{n < \omega}
\textbf{2}^{\D^n}$, with $P$ a set of predicate symbols of the
language and $\textbf{2} = \{t,f\}$ is the set of truth values (true
and false, respectively), such that for any world $w \in \mathbb{W}$
and a predicate symbol $p \in P$, we obtain a function
$V(w,p):\D^{arity(p)} \rightarrow \textbf{2}$,
 which defines the extension $[p] = \{\textbf{a} | \textbf{a} \in \D^{arity(p)}~$ and $~ V(w,p)(\textbf{a}) = t\}$
 of this predicate $p$ in the world $w$.\\
 The extension of a formula $A$, w.r.t. a model $\M$, a
 world $w \in \mathbb{W}$ and an assignment $g:Var \rightarrow \D$ is denoted by $~[A]^{\M, w,
 g}$ or by $~[A/g]^{\M, w}$ where $A/g$ denotes the formula obtained from $A$ by assigning (with $g$) the values to all its free variables.
 Thus, if $p \in F\bigcup P$ then for a given world $w \in \mathbb{W}$ and
 the assignment function for variables $g$, $~[p]^{\M, w, g} =
 V(w,p):\D^{arity(p)}\rightarrow \textbf{2}$, that is, for any set of terms $t_1,..,t_n$, where $n$ is
 the arity of $p$, we have $~[p(t_1,..,t_n)]^{\M, w, g} =
 V(w,p)([t_1]^{\M, w, g},..,[t_n]^{\M, w, g})~ \in \textbf{2}$.\\
% with terms defined  by:\\
% 1. All variables $x \in Var$ and the constants $d \in \D$ are terms;\\
% 2. If $f \in F$ is a function symbol of arity $n$, and
% $t_1,..,t_n$ are terms, then a functional form $f(t_1,..,t_n,)$ is a  term.\\
 For any formula $A$,  $~~\M \vDash_{w,g} A~$ is equivalent to $~ [A]^{\M, w, g}
 = t$, means 'A is true in the world $w$ of a model $\M$ for
 assignment $g$'.\\
  The additional semantic
rules relative to the modal operators $\square$ and
$\lozenge$ are as follows:\\
  $\M \vDash_{w,~g} \square A~~~~~~$ iff $~~~~\M \vDash_{w',~g} A~~$
  for every $w'$ in $\mathbb{W}$ such that $w \R w'$.\\
  $\M \vDash_{w,~g} \lozenge A~~~~~~$ iff there exists a $w'$
  in $\mathbb{W}$ such that $w \R w'$ and $~~\M \vDash_{w',~g} A~$.\\
A formula $A$ is said to be \emph{true in a model} $\M~~$  if $~\M
\vDash_{w,~g} A~~$ for each  $g~$ and  $w\in \mathbb{W}$ . A formula
is said to be \emph{valid} if it is true in
each model.\\
 Montague defined the \emph{intension} of a formula $A$
 as follows:\\
 $~[A]_{in}^{\M, g} =_{def} \{ w \mapsto [A]^{\M, w,
 g} ~|~ w \in \mathbb{W}\}$,\\
 i.e., as graph of the function $~[A]_{in}^{\M, g}:\mathbb{W} \rightarrow \bigcup_{w \in \W_N} [A]^{\M, w,
 g}$.  \\
One thing that should be immediately clear is that intensions are
more general then extensions: if the intension of an expression is
given, one can determine its extension with respect to a particular
world but not viceversa, i.e., $~[A]^{\M, w,
 g} = [A]_{in}^{\M, g}(w)$.\\
 In particular, if $c$ is a non-logical constant (individual
 constant or predicate symbol), the definition of the extension of
 $c$ is, $~[c]^{\M, w, g} =_{def} V(w,c)$. Hence, the intensions
 of the non-logical constants are the following functions:
$~~~~[c]_{in}^{\M, g}:\mathbb{W} \rightarrow \bigcup_{w \in \mathbb{W}} V(w,c)$.\\
The extension of variable is supplied by the value assignment $g$
only, and thus does not differ from one world to the other; if $x$
is a variable we have $~~~[x]_{in}^{\M, g} = g(x)$.\\
Thus the intension of a variable will be a constant function on
worlds which corresponds to its extension. Finally, the connection
between Bealer's non-reductionistic  and Montague's possible world
approach to intensional logic can be given by the isomorphism (its
meaning is that basically we can use the extensionalization
functions in the place of Montague's possible worlds):
%\begin{center}
$~~~~~~~~~~~\F:\mathbb{W} \simeq \E$,
% \end{center}
\\  where  $\E$ is a set of possible extensionalization functions
which can be considerd as possible worlds (up to the previous
isomorphism): Each extensionalization function $h \in \E$ assigns to
the  intensional elements of $\D$ an appropriate extension as
follows:\\
 for each proposition $u \in D_0$, $h(u) \in \textbf{2} = \{f,t\}$ is its
 extension (true or false value); for each n-ary
 relation-in-intension $u \in D_n$, $h(u)$ is a subset of $\D^n$
 (n-th Cartesian product of $\D$); in the case of particulars $u \in
 D_{-1}$, $h(u) = u$.
We require that operations $conj, disj$ and $neg$ in this
intensional algebra behave in the expected way with respect to each
extensionalization function (for example, for all $u \in D_0$,
$h(neg(u)) = t$ iff $h(u) = f$, etc..), that is
\begin{center}
 $h = h_{-1} + h_0 + \sum_{i\geq 1}h_i:\sum_{i
\geq -1}D_i \longrightarrow D_{-1} + \textbf{2} + \sum_{i\geq
1}\P(D^i)$
\end{center}
 where $h_{-1} = id:D_{-1} \rightarrow D_{-1}$
is identity, $h_0:D_0 \rightarrow \textbf{2}$ assigns truth values
in $\textbf{2} = \{f,t\}$, to all propositions, and $h_i:D_i
\rightarrow \P(D^i)$, $i\geq 1$, assigns extension to all
relations-in-intension, where $\P$ is the powerset operator. Thus,
intensions can be seen as \emph{names} of abstract or concrete
entities, while extensions correspond to various rules that these
entities play in different worlds. Among the possible functions in
$\E$ there is
 a distinguished function $\Bbbk$ which is to be thought as the
 \emph{actual}  extensionalization function: it tells us the
 extension of the intensional elements in $\D$ in the current actual
 world.
 % $\vspace*{-3mm}$
 In what follows we will use the join operator $\bowtie$, such that
for any two relations $r_1, r_2$ their join is defined by: $~~r_1
\bowtie r_2 =
\{(\textbf{a},\textbf{c},\textbf{b})~|~(\textbf{a},\textbf{c}) \in
r_1$ and $(\textbf{c},\textbf{b}) \in r_2 \}$, $~~$ where
$\textbf{a},\textbf{c},\textbf{b}$ are tuples (also empty) of
constants, so that $~~~r_1 \bowtie \{\} = \{\}$ and $~~~r_1 \bowtie
\{<>\} = r_1$.
 \begin{definition} (SEMANTICS): \label{def:Semantics}
The operations of the algebra $~Alg_{int}$ must satisfy the
following conditions, for any $h \in \E$, with $f = \{\}, t = \{<>\}$, and $u_1,..,u_i \in \D$:\\
1. $~~~h(conj(u,v)) = h(u) \bowtie h(v)$, $~~$ for $u,v \in D_I$.\\
%1.1 $~~~<x_1,..,x_i> \in h(conj(u,v))~~~$ iff
%\\$~~~~~~~~~~~<x_1,..,x_i> \in
%h(u)\bigcap h(v)$, $~~$ for $u,v \in D_i, i\geq 1$.\\
%1.2 $~~~h(conj(u,v)) = t~~~$ iff $~~~h(u) = h(v) = t$, $~~$ for $u,v \in D_0$.\\
2.1 $~~~h(neg(u)) = t~~~$ iff $~~~h(u) = f$, $~~$ for $u \in D_0$.\\
2.2 $~~~<u_1,..,u_i> \in h(neg(u))~~~$ iff $~~~<u_1,..,u_i> \notin
h(u)$, $~~$ for $u \in D_i, i\geq 1$.\\
%3. $~~~disj (u,v) = neg(conj(neg(u), neg(v)))$, $~~$ for  $u,v \in D_i, i\geq 0$.\\
%4. $~~~impl(u , v) = disj(neg(u) , v)$, $~~$ for $u,v \in D_i, i\geq 0$.\\
%5.1 $~~~h(exist(u)) = t~~~$ iff $~~~h(u) = t$, $~~$ for $u \in D_0$.\\
%5.2 $~~~h(exist(u)) = t~~~$ iff $~~~\exists x_1(x_1 \in h(u))$, $~~$ for $u \in D_1$.\\\
%5.3 $~~~<x_1,..,x_{i-1}> \in h(exist(u))~~~$ iff $~~~(\exists x_i)(<x_1,..,x_{i-1},x_i> \in h(u))$,\\ $~~$ for $u \in D_i, i\geq 2$.\\
3.1 $~~~h(pred(u,u_1)) = t~~~$ iff $~~~u_1 \in h(u)$, $~~$ for $u \in D_1$.\\
3.2 $~~~<u_1,..,u_{i-1}> \in h(pred(u,u_i))~~~$ iff
\\$~~~~~~~~~~~<u_1,..,u_{i-1},u_i> \in
h(u)$, $~~$ for $u \in D_i, i\geq 2$.
% $\vspace*{-2mm}$
\end{definition}
Notice that this definition for the \emph{semantics} of the
conjunction operation is different from the original work in \cite{Beal79} where\\
1.1 $~~~<u_1,..,u_i> \in h(conj(u,v))~~~$ iff
\\$~~~~~~~~~~~<u_1,..,u_i> \in
h(u)\bigcap h(v)$, $~~$ for $u,v \in D_i, i\geq 1$.\\
1.2 $~~~h(conj(u,v)) = t~~~$ iff $~~~h(u) = h(v) = t$, $~~$ for $u,v \in D_0$.\\
Once one has found a method for specifying the denotations of
singular terms of $\L_{\omega}$ (take in consideration the
particularity of abstracted terms), the Tarski-style definitions of
truth and validity for  $\L_{\omega}$ may be given in the customary
way.  An \emph{intensional interpretation} $I$ \cite{Beal82}
%, for a given assignment $g:Var \rightarrow \D$ for variables of $\L_{\omega}$,
% is a homomorphism $I:Pr\rightarrow \D$, that
 maps each i-ary predicate letter  of
$\L_{\omega}$ to i-ary relations-in-intention in $D_i$. It can be
extended to all formulae in usual way. What is being sougth
specifically is a method for characterizing the denotations of
singular terms of $\L_{\omega}$  in such a way  that a given
singular term $\langle A \rangle_{x_1...x_m}$ will denote an
appropriate property, relation, or proposition, depending on the
value of $m$. We denote by $A_{BS}$ the set of intensional abstracts
(terms, so that $A_{BS} \subset \L_{\omega}$. Thus, the mapping
$den:A_{BS} \rightarrow \D$ given in the original version of Bealer
\cite{Beal82} will be called \emph{denotation}, such that the
denotation of $\langle A\rangle $
 is equal to the meaning of a proposition $A$, that is, $~~den(\langle A\rangle) =
I(A)\in D_0$.  In the case when $A$ is an atom $F^m(x_1,..,x_m)$
then $den \langle F^m(x_1,..,x_m)\rangle_{x_1,..,x_m} = I(F^m) \in
D_m$. The denotation of a more complex abstract $\langle
A\rangle_\alpha$ is defined in terms of the denotation(s) of the
relevant syntactically simpler abstract(s) \cite{Beal82}.\\
For
example $I(A(x) \wedge B(x)) = conj(I(A(x)), I(B(x)))$, $I(\neg p) =
neg(I(p))$.
% where $exist$ is an intensional algebraic operator, analogue of the existential operator in polyadic algebras (polyadic
%algebra is an extensional model of FOL).
 A sentence $A$ is true
relative to $I$ and the intensional algebra, iff its \emph{actual}
extention is equal to $t$, that is, $~Tr(\langle A \rangle)~$ iff
$~\Bbbk(I(A)) = t$,
where $Tr$ is unary predicate for true sentences.\\
For the predicate calculus with individual constants (variables with
fixed assignment, proper names, and intensional abstracts) we
introduced an additional binary algebraic operation $pred$ (singular
predication, or \emph{membership} relation), such that for any two
$u,v \in \D$, for any extensionalization function $h$  holds
$~h(pred(u,v)) = t~$ iff $~v \in h(u)$. So we are able to assign
appropriate intensional value (propositional meaning) to a ground
atom $A(c) \in \L_{\omega}$ with individual constant $c$.\\
 That is,
$I(A(c)) = pred(I(A(x)), I(c)) $ is an expression in this
intensional algebra with $I(A(x)) \in D_1$ and $I(c) \in D_{-1}$. So
that $h(I(A(c))) = h(pred(I(A(x)), I(c)) = t$ iff $I(c) \in
h(I(A(x)))$. That is, in the 'world' $h$, $A(c)$ is true (that is,
the extension of the propositional meaning of $A(c)$ is equal to
$t$) iff the interpretation of $c$ is in the extension of the
interpretation of the predicate $A(x)$. Or, for example, for a given
formula with intensional abstract,
 $B(\langle A(x,y) \rangle_{x,y}) \in \L_{\omega}$, we have that  $h(I(B(\langle A(x,y)
\rangle_{x,y})))\\ = h(pred(I(B(z)), den(\langle A(x,y)
\rangle_{x,y}))) = t~$ iff $~den(\langle A(x,y) \rangle_{x,y}) \in
h(I(B(z)))$, where $~~I(B(z)) \in D_1$ and $~~den(\langle A(x,y)
\rangle_{x,y}) \in D_2$.
% $B(\langle A(c) \rangle) \in \L_{\omega}$, we have that  $h(I(B(\langle A(c) \rangle))) =
%h(pred(I(B(z)), den(\langle A(c) \rangle))) = t~$ iff
%$~den(\langle A(c) \rangle) \in h(I(B(z)))$, where $I(B(z)) \in
%D_1$ and $den(\langle A(c) \rangle) \in D_0$.
 \\
 We can connect $\E$
 with a possible-world semantics, where $w_0 = \F^{-1}(\Bbbk)$ denotes the actual
 world in which intensional elements have the extensions defined by
 $\Bbbk$. Such a correspondence, not present in original intensional
 theory \cite{Beal93}, is a natural identification of
 intensional logics with modal Kripke based logics.
 %$\vspace*{-3mm}$
\begin{definition} (Model): \label{def:Semant} A model for the
intensional FOL is the S5 Kripke structure $\M_{int} = (\mathbb{W},
\R, \D, V)$ where $\mathbb{W} = \{\F^{-1}(h)~|~h \in \E\}$, $~~\R =
\mathbb{W} \times \mathbb{W}$. \\Intensional identity "$=$" between
ground intensional terms $\langle A \rangle_{\alpha}/g$ and $\langle
B \rangle_{\alpha}/g$, where all free variables (not in $\alpha$)
are instantiated by $g \in \D^{Var}$, is defined as
follows:\\
%\begin{center}
$\langle A \rangle_{\alpha}/g =  \langle B \rangle_{\alpha}/g~~~~~$
iff $~~~~~\square ( A^{\alpha}/g \equiv B^{\alpha}/g)$\\
%\end{center}
where  $A^{\alpha}/g$ denotes the logic formula where all free
variables not in $\alpha$ are instantiated by the assignment $g$. If
$\alpha$ are all free variables in $A$ then instead of $\square
A^{\alpha}/g$ we write simply $\square A$.
% and $~\D = D_{-1} \bigcup D_0 \bigcup D_1 ...\bigcup D_n ...$ domain in Definition \ref{def:Syntax}.
The symbol $\square$ is the universal "necessity" S5 modal operator.
 %$\vspace*{-4mm}$
 \end{definition}
 Let $A/g$ denote the ground formula obtained from a formula with
 free variables $A$ and an assignment $g:Var \rightarrow \D$. Then the
satisfaction relation $\models$ for this Kripke semantics is defined
by, $~~\M \models_{w,g} A~~$ iff $~~\F(w)(I(A/g)) = t$.\\
Remark: this semantics is equivalent to the algebraic semantics for
$\L_{\omega}$ in \cite{Beal79}
%for the case of the  conception
where intensional entities are considered to be \emph{identical} if
and only if they are \emph{necessarily equivalent}. Intensional
identity is much stronger that the standard \emph{extensional
equality}  in the actual world, just because requires the
extensional equality in \emph{all} possible worlds, in fact, if
$\langle A \rangle_{\alpha}/g =  \langle B \rangle_{\alpha}/g~$ then
$h(den(\langle A \rangle_{\alpha}/g)) =  h(den(\langle B
\rangle_{\alpha}/g))~$ for all extensionalization functions $h \in
\E$ (that is possible worlds $\F^{-1}(h) \in \mathbb{W}$). But we
can have the extensional equality in the possible world $w =
\F^{-1}(h)$, while $den(\langle A \rangle_{\alpha}/g) \neq
den(\langle B \rangle_{\alpha}/g)~$, that is, when $A$ and $B$ are
not intensionally equal, so that each intensional identity class of
elements is the subset of
the extensional equivalence class.\\
Moreover, for this intensional FOL holds the soundness an
completeness: For all formulae $A$ in $\L_{\omega}$, $A$ is valid if
and only if $A$ is a theorem of this First-order S5 modal logic with
intensional equality
\cite{Beal79}.\\
It is easy to verify that the intensional equality means that in
every possible world $w \in \mathbb{W}$ the intensional entities $A$
and $B$ have the same extensions (as in Montague's approach),
moreover:
\begin{propo} (\emph{Bealer-Montague connection}): For any
intensional entity $\langle A/g\rangle$ its extension in a possible
world $w \in \mathbb{W}$ is equal to $~~~~~~\F(w)(den(\langle A/g
\rangle)) = ~[A]_{in}^{\M, g}(w)$.
%,\\ where $den$ means 'denotation', $den(\langle A_i(\textbf{x}) \rangle_{\alpha}) \in \D$.
%$\vspace*{-4mm}$
\end{propo}
\textbf{Proof:} Directly from the definition of the identification
of a possible world $w $ of Montague's approach with the extensional
function $h = \F(w) \in \E$ in the Bealer's approach, where
$~[A]_{in}^{\M, g}$ is the "functional" intension of Montague, and
$\langle A \rangle$ is an intensional term of Bealer's logic.
\\$\square$
% \\Now we can introduce the new intensional equivalence relation between intensional entities:
%$\vspace*{-3mm}$
\begin{definition}(Intensional Equivalence $~\approx~$): \label{def:equival}
 The  intensional ground terms $\langle A \rangle_{\alpha}/g$ and $\langle B
 \rangle_{\alpha}/g$,
where the assignment $g$ is applied only to free variables not in
$\alpha$,  are intensionally  equivalent,
%\begin{center}
$~~~~~~~~\langle A \rangle_{\alpha}/g \approx \langle B
\rangle_{\alpha}/g~~~~$ iff $~~~~\lozenge A^{\alpha}/g \equiv
\lozenge
B^{\alpha}/g$, $~~~$\\
%\end{center}
where $\lozenge = \neg \square \neg$, and $A^{\alpha}/g$ denotes the
logic formula where all free variables not in $\alpha$ are
instantiated by the asignment $g \in \D^{Var}$. If $\alpha$ are all free variables in $A$
then instead of $\lozenge A^{\alpha}/g$ we write simply $\lozenge A$.\\
% is the existential modal operator of a Kripke structure in Definition \ref{def:Semant}.\\
This equivalence defines the \textsc{quotient} algebra
$~Alg_{int}/_{\approx}$ for a quotient-intensional FOL
$\L_{\omega}/_{\approx}$, as follows:\\
Given an intensional logic $\L_{\omega}$ with  a basic, user
defined, set of intensional equivalences $S_{eq}$, and its deductive
inference relation $ \vdash_{in}$ of the S5 modal logic with
intensional equality theory, then, for any intensional entity
$\langle A(\textbf{x}) \rangle_{\textbf{x}}$, where $\textbf{x} =
<x_1,..,x_k>$ is a tuple of free variables in $A$, we obtain an
intensional-equivalence class \\$\C = \{\langle A_i(\textbf{x})
\rangle_{\textbf{x}}~|~ A_i(\textbf{x}) \in \L_{\omega}~$, such that
$~~\L_{\omega},S_{eq}~\vdash_{in}~  \langle A_i(\textbf{x})
\rangle_{\textbf{x}} \approx \langle A(\textbf{x})
\rangle_{\textbf{x}}\}$.\\If we denote by $\langle A(\textbf{x})
\rangle_{\textbf{x}} \in ~Alg_{int}/_{\approx}$ the quotient
intensional entity for this equivalence class,
% Let $\langle A(\textbf{x})
%\rangle_{\alpha}$, where $\textbf{x} = \{x_1,..,x_k\}$ are free
%variables, be the quotient intensional entity for the equivalence
%class $\C = \{\langle A_1(\textbf{x}) \rangle_{\alpha},...,
%\langle A_m(\textbf{x}) \rangle_{\alpha}\}$ then its
its extension in a world $w$ is defined by\\
$\F(w)(den(\langle A(\textbf{x}) \rangle_{\textbf{x}})) =
\{\textbf{t}\in
\D^k~|~A_i(\textbf{t})$ is true in w,  $~ A_i(\textbf{x})\in \C\}$\\
$~~~~~~~~~~~~~~~~~~~~~~~~~~~~~~~= \bigcup_{1\leq i \leq m}
\F(w)(den(\langle A_i(\textbf{x}) \rangle_{\textbf{x}}))$.
%$\vspace*{-4mm}$
\end{definition}

 This definition of equivalence relation is the flat-accumulation
case presented in \cite{Majk03s,Majk04ph}: if the first predicate is
true in some world then the second must be true in some world also,
and vice versa. Each equality is also intensional equivalence, but
not vice versa.\\
Let the logic modal formula $\lozenge A^{\alpha}/g$, where the
assignment $g$ is applied only to free variables of a formula $A$
not in the list of variables in $\alpha = <x_1,...,x_n>$, $n \geq
1$, represents an n-ary intensional concept such that $I(\lozenge
A^{\alpha}/g) \in D_n$ and $I(A^{\alpha}/g) = den(\langle A
\rangle_{\alpha}/g) \in D_n$. Then the extension of this n-ary
concept is equal to (here the mapping $posib:D_i\rightarrow D_i$ for
each $i \geq 1$
is a new operation of the intensional algebra $~Alg_{int}$ in Definition \ref{def:Syntax}):\\
$ h(I(\lozenge A^{\alpha}/g)) = h(posib(I(A^{\alpha}/g))) =\\
 = \{<g'(x_1),...,g'(x_n)>~|~\M
\models_{w,g'} \lozenge A^{\alpha}/g~$ and $g' \in \D^{Var}\}$\\
$ = \{ <g'(x_1),...,g'(x_n)>~|~\exists w_1 ((w,w_1) \in \R$ and $\M
\models_{w_1,g'}  A^{\alpha}/g)~\}$\\
$ = \{<g'(x_1),...,g'(x_n)>~|~\exists h_1 (  \M
\models_{\F^{-1}(w_1),g'} A^{\alpha}/g)~\}$\\
$= \bigcup_{h_1 \in ~\E} h_1(I(A^{\alpha}/g)) = \bigcup_{h_1 \in ~\E} h_1(den(\langle A \rangle_{\alpha}/g))$.\\
It is easy to verify that the n-ary concepts $A^{\alpha}/g$ and
$\lozenge A^{\alpha}/g$ are intensionally equivalent (i.e., it holds
that $\lozenge A^{\alpha}/g \equiv \lozenge \lozenge A^{\alpha}/g
$), the first  is denominated contingent-world entity because its
extension depends on a particular world, while the second in
denominated real-world entity because its extension is constant,
i.e. fixed for every world, as explained in \cite{Majk04ph}. That
is, the concept $\lozenge A^{\alpha}/g$ is a \emph{built-in} (or
rigid) concept whose extension does not depend on possible worlds,
and can be considered as representative element (with maximal
extension) for each class of intensionally equivalent concepts.\\
Analogously, the "necessity" intensional operator
$necess:D_i\rightarrow D_i$ for each $i \geq 1$
is a new operation of the intensional algebra $~Alg_{int}$ in Definition \ref{def:Syntax} we have that:\\
$ h(I(\square A^{\alpha}/g)) = h(necess(I(A^{\alpha}/g))) =\\
 = \{<g'(x_1),...,g'(x_n)>~|~\M
\models_{w,g'} \square A^{\alpha}/g~$ and $g' \in \D^{Var}\}$\\
$ = \{ <g'(x_1),...,g'(x_n)>~|~\forall w_1 ((w,w_1) \in \R$ implies
$\M
\models_{w_1,g'}  A^{\alpha}/g)~\}$\\
$= \bigcap_{h_1 \in ~\E} h_1(I(A^{\alpha}/g)) = \bigcap_{h_1 \in ~\E} h_1(den(\langle A \rangle_{\alpha}/g))$.\\
Consequently, the concept $\square A^{\alpha}/g$ is a
\emph{built-in} (or rigid) concept as well, whose extension does not
depend on possible worlds. \\For example, for two intensionally
equal ground terms $\langle A \rangle_{\alpha}/g$ and $\langle B
\rangle_{\alpha}/g$, we have that $~~~h(I(\square ( A^{\alpha}/g
\equiv B^{\alpha}/g))) =
\D^n.$\\
 In what concerns this paper we will consider
\emph{only} the actual world $w_0 = \F^{-1}(\Bbbk)$. Moreover, the
set of \emph{basic} intensional equivalences are designed by users,
and we will not verify if they satisfy the modal formula used to
define the intensional equivalence: the definition above has a
theoretical interest, but useful to understand the meaning of the
intensional equivalence and the "omniscient" inference relation $
\vdash_{in}$, able to deduce all other intensional equivalences from
a given basic set.
\begin{propo} \label{prop:equival} Let $C(\textbf{x})$ be a logic
formula defined from built-in predicates (ex, $\leq, \geq$, etc.),
then $~\langle A(\textbf{x}) \rangle_{\textbf{x}} \approx \langle
B(\textbf{x}) \rangle_{\textbf{x}}~~~$ implies $~~~\langle
A(\textbf{x}) \wedge C(\textbf{x}) \rangle_{\textbf{x}} \approx
\langle B(\textbf{x}) \wedge C(\textbf{x}) \rangle_{\textbf{x}}$
%$\vspace*{-4mm}$
\end{propo}
\textbf{Proof:} immediately from the fact that a built-in formulae
$C(\textbf{x})$ has constant extension in any possible world in
$\mathbb{W}$.  \\$\square$\\
The quotient intensional FOL $\L_{\omega}/_{\approx}$ (its algebraic
counterpart is a Lindenbaum-Tarski algebra) is fundamental for
\emph{query answering }in intensional P2P database mapping systems:
given a query $q(\textbf{x})$ over a peer $P_i$, the answer to this
query is defined as the extension of the denotation of the
intensional concept $\langle q(\textbf{x}) \rangle_{\textbf{x}}$,
% where $K_i$ is the epistemic modal "know" operator of a peer database $P_i$,
 in
the intensional P2P logic $\L_{\omega}/_{\approx}$.

\section{Embedding of P2P database systems into intensional FOL}
    The formal semantic framework for P2P database systems, presented also in ~\cite{Majk03s}
 as a hybrid modal logic, in this paper will be defined as
 quotient (by intensional equivalence) intensional FOL.\\
 We will consider \emph{only} the actual world $w_0
= \F^{-1}(\Bbbk)$, correspondent to the extensionalization function
$\Bbbk$ of the quotient intensional FOL $\L_{\omega}/_{\approx}$:
the actual world for $\L_{\omega}/_{\approx}$ corresponds to the
actual extension of peer databases. When an user defines a
conjunctive query $q(\textbf{x})$ over an ontology $\O_i$ of a peer
database $Pi$, the answer to this query is computed in this actual
world $w_0$, that is in the actual extension of all peer databases
in a P2P network $\N = \{P_i~|~ 1\leq i \leq N \}$.
\begin{definition} \label{def:P2Plogic} Let $\N = \{P_i~|~ 1\leq i \leq N \}$ be a P2P database system.
 The  intensional FOL  $~\L_{\omega}~$ for a query answering
in a P2P network $\N$ is composed by: \\
1. The set of basic intensional entities is a disjoint union of
entities of peers $~S_I = \biguplus_{1\leq i \leq N} \{
r(\textbf{y})
 ~|~ r(\textbf{y}) \in \O_i \}$.
 %All other intensional entities are obtained from this basic set by recursive
%application of operators of the intensional algebra $~Alg_{int} = <\D, conj, disj, impl, neg, \tau >~$ (Definition \ref{def:Syntax}).
The intensional interpretation of the set of all intensional
entities define the
domains $D_n, n \geq 1$;\\
 2. The extensional part of a domain, $D_{-1}$, corresponds to the
 disjoint union of domains of peer databases.
 %, extended by logic values  in $\textbf{2} = \{f,t\}$;
 The intension-in-proposition part, $D_0$, is defined by disjoint union of peer's Herbrand bases.\\
 3. The basic set of the equivalence relation $~\approx~$
 is defined as disjoint union for each peer $P_i$ as follows ($\textbf{x}$ is a tuple of variables of queries):\\
  if $~~~~(q^{ij}_{1k}(\textbf{x}), q^{ij}_{2k}(\textbf{x})) \in
 \M^{ij}~~$, then $~~~~\langle q^{ij}_{1k}(\textbf{x}) \rangle_{\textbf{x}} ~\approx ~\langle q^{ij}_{2k}
 (\textbf{x}) \rangle_{\textbf{x}}$.\\
  The \textsc{complete} P2P answer to
a conjunctive query $q(\textbf{x})$ over a peer $P_i$ is equal to
the extension of the quotient-intensional concept $\langle
q(\textbf{x}) \rangle_{\textbf{x}}$, whose equivalence class is
determined by the deductive omniscient closure of $~\vdash_{in}$, in
the quotient intensional P2P logic $\L_{\omega}/_{\approx}$.
\end{definition}
Notice that in this embedding of  a P2P system, into the intensional
FOL  $~\L_{\omega}~$, we do not use any existential quantifier, so
that the intensional algebra in Definition \ref{def:Syntax} is
sufficient for a P2P query answering.
%\textbf{Remark:} Notice that for each conjunctive query $q \equiv
%r_1 \wedge ...\wedge r_n$ over a peer $P_i$, so that  $r_1, ..., r_n
%\in \O_i$, from the fact that each peer is a S5 epistemic
%\emph{normal} modal logic, we have that $K_i q \equiv K_i(r_1 \wedge
%...\wedge r_n) = K_i r_1 \wedge ...\wedge K_ir_n$, that is
%decomposable into conjunction of basic intensional entities in
%$\L_{\omega}/_{\approx}$, so that the definition above for
%intensional equivalence relation and user conjunctive queries is complete.\\
We need to make complete the model of the intensional logic
$\L_{\omega}/_{\approx}$ by defining its
extensionalization function $\Bbbk$ for the actual world $w_0$.\\
For this aim we will consider as actual world $w_0$, of the
intensional
logic, the actual \emph{extensional} FOL multi-modal P2P database system:\\
  What we obtain is a two-level modal framework: the higher, or
\emph{P2P query answering}, level is the Bealer's intensional  logic
(without quantifiers) with S5 Montague's possible-worlds
$\mathbb{W}$ modal structure, where $w_0 \in \mathbb{W}$ is actual
world for a P2P system; the lower, "computational", level is the
extensional FOL  multi-modal epistemic logic (with existential
quantifiers also)
%, where the set of worlds $\W$ is an extension of the union of sets of preferred Herbrand models
 of each peer
database.
%Consequently, each Bealer/Montague's possible world contains a particular set of low-level worlds of the extensional FOL
%multi-modal epistemic P2P system.
We can see this "computational" level as a sophisticated wrapper
(based on the FOL logic of a Data Integration System which is
encapsulated into a peer as an Abstract Data Type (ADT)
\cite{Majk06J}). Each peer is considered as an \emph{independent}
(from other peers) sophisticated wrapper, which extracts the exact
extension (of \emph{only known} facts) for all predicates used in
upper intensional P2P query answering logic layer. The extension of
each peer database models the extensionalization function for
intensional FOL with intensional equivalence, used for intensional
embedding of a P2P database system given by Definition
\ref{def:P2Plogic}. The details of the computation of this
extensionalization function $\Bbbk$,
for the actual Montague's world $w_0$, can be found in \cite{Majk08f,Majk08in}.\\
%2 - Here we will use the standard \emph{extensional} multi-modal
%logic framework, of this lower "computational" level of each
%independent peer. We will differentiate the accessibility relation
%between peers (network's (or \emph{global}) conceptual level) from
%the accessibility relations which model the epistemic query
%answering semantics of \emph{local} peers: this multi-modal logic
%will define the extensionalization function $\Bbbk$ for the
%actual world $w_0$.\\
\textbf{Remark}: This is very important observation. What we
obtained is relatively simple intensional logic without quantifiers,
with only a subset of predicates used in the global schema of peer
databases with  the set of views (virtual predicates)  defined for
intensional mapping between peers. The extension of these predicates
is wrapped by ADT of each peer independently. The logic
specification for these sophisticated wrappers can be obtained by
using the epistemic FOL logic \cite{Majk06J} of each single peer
database.\\
%Here a peer is considered as a Data Integration System with Global-As-View (GAV) mappings between its source and its global
%database schemas \cite{Lenz02}, and with integrity constraints over
%global schema also, which possibly can use the existential quantifiers.\\
Each peer database architecture uses the strong (extensional)
semantic Global-As-View (GAV) mapping, based on views, \emph{inside}
each peer database, as in standard Data Integration Systems
\cite{LeLR02,Majk05p}, with the possibility to use also the logic
negation \cite{Majkw04}. The weak (intensional) semantic mapping
based on views, is used for the mapping \emph{between} the peers.
This architecture takes advantage of both semantical approaches:
extensional for building independent peer databases (a development
of any particular per database can be done by a compact group of
developers, dedicated to developing and to maintaining  its
functionalities), with intensional, robust and non invasive, mapping
between peers, based on beliefs of developers of one peer about the
intensionally equivalent knowledge of other peers (which are not
under their control).\\
The actual world $w_0$, with correspondent extensionalization
function $\Bbbk = \F(w_0)$, is represented as an \emph{extensional}
FOL multi-modal logic theory for a P2P database system, composed by
a number of peers $\{P_i~|~ 1\leq i \leq N \}$, defined
%~\cite{Majk03s} in this hybrid language
as follows:
%\\ Thus, the hole logic theory for
%peers and their mappings can be defined by this multi-modal logic
%for P2P database systems. The frame and the model for this
%multi-modal  translation of a P2P database system can be given now
%by the following definition.
\begin{definition} \label{def:kripke}
 We consider a model $\M$, for the extensional multi-modal
logic translation of a P2P database system composed by N peers,  a
four-tuple $(\W,\{\R_i\},\D,\V)$ , where:
\begin{itemize}
  \item The set of points is a disjoint union $~\W = \sum_{1 \leq i \leq N}
(\W_i \bigcup \{P_i\})$, with $\W_i =  Mod(P_i)$,  where:\\
  1.  Each point  $P_i$ is considered as a FOL theory with incomplete information,
  composed by an extensional (ground atoms/facts) and, possibly, an intensional  part (logic formulae with variables).\\
%  For example, each point $\A \in \W$ can be considered as a logic
%  (or more simple, Datalog) program which defines some peer database.\\
  2. For each peer database $P_i$, the set of points $~ ~\W_i =  Mod(P_i), ~ ~1 \leq i \leq N$ is the set of all preferred Herbrand models
  of   such  peer database. Each $w \in Mod(P_i)$ can be  seen as a logical theory also, composed by only
ground terms (only extensional part).
%\\A  peer database $\A$ with all its minimal models $\A_{ik}$ is a disjoint partition in $\W$.
\item $\R_0$ is a binary accessibility relation  between peers, such that $(P_i,P_j) \in \R_0$ iff
a mapping exists from  peer $P_i$ to  peer $P_j$. Then we close this
relation for its reflexivity and transitivity properties.
    \item  $~\R_i = \{P_i\} \times \W_i,~ 1\leq i \leq N$  is a binary
accessibility relation for a i-th peer universal modal operator
$K_i$, so that, for a given view $q(\textbf{x})$ over a peer $P_i$,
and assignment $g$, $~~~\M \models_{P_i,g}K_iq(\textbf{x})~~~$ iff
$~~\forall w ((P_i, w) \in \R_i$ implies $~\M
\models_{w,g}q(\textbf{x})).$
%    on a $\W_i$ for the S5 epistemic logic of each peer $P_i$.
%   Thus, it is reflexive symmetric and transitive relation, used for
%the "know" universal modal operator $K_i$ of the peer $P_i$.
  \item $\V$ is a function which assigns to each pair consisting
  of an n-place predicate constant $r$ and of an element $w \in
  \W$ a function $\V(r,w)$ from $\D^n$ to $\{ 1, 0\}$.
%  \item $\V_N$ is a function which assigns to each nominal $i$, a    point $P_i \in \W$.
\end{itemize}
So, the extensionalization function $\Bbbk = \F(w_0)$
%in this actual world $w_0 \in \mathbb{W}$
for basic intensional entities of the intensional P2P logic
$\L_{\omega}/_{\approx}$, is defined as follows: for any  $~\langle
r(\textbf{y}) \rangle$, where $~r$ is an n-ary (virtual) predicate
of a peer $P_i$, and $\textbf{y}, \textbf{c}$ are n-tuples of
variables and constants in $\D$ respectively, we define
\begin{itemize}
  \item for any n-ary relation-in-intension
$~~den(\langle  r(\textbf{y}) \rangle_{\textbf{y}}) \in D_n$, $n
\geq 1$, \\$~~\Bbbk(den(\langle  r(\textbf{y})
\rangle_{\textbf{y}})) = \{g(\textbf{y}) ~|~\M \models_{P_i,g}
K_ir(\textbf{y}), ~$ and assignment $g:Var \rightarrow\D \}$.
  \item for intensional propositions $ den(\langle  r(\textbf{c})
\rangle)$ in $D_0$, \\ $\Bbbk(den(\langle  r(\textbf{c}) \rangle)) =
t~~~$ if $~~\M \models_{P_i} K_ir(\textbf{c}) ~~$; $~~f~~$,
otherwise.
\end{itemize}
 \end{definition}
 In this way the binary  relation of each partition (peer database), $\R_i, i \geq 1$,
  models the \emph{local} universal epistemic modal operator $K_i$ for each peer
database. In fact it holds that $~~~\M
\models_{P_i,g}K_iq(\textbf{x})~~~$ iff $~~\forall w \in Mod(P_i)
(~\M \models_{w,g}q(\textbf{x}))$, i.e., $K_iq(g(\textbf{x}))$ is
true iff $q(g(\textbf{x}))$ is true in \emph{all} preferred models
of a peer $P_i$. The binary relation $\R_0$, instead, models the
\emph{global epistemic} P2P modal operator $\K$ \cite{Majk08in} in
this extensional multi-modal logic.\\
\emph{Context-dependent} query answering: notice, that the answer to
any query depends on the topology of the P2P network, that is, it
depends on the peer's accessibility relation $\R_0$, so that for
equivalent queries, but formalized over different peers we will
generally obtain different answers. Now we are able to synthesize
the definition of intensionally equivalent views used for mappings
between peers, in this two-leveled Kripke  model framework:
\begin{definition} (Intensional FOL  for P2P systems): \label{def:SemantP2P} A two-level Kripke model for the
intensional FOL  of a P2P database system $\N$, given in Definition
\ref{def:P2Plogic}, is the S5 Kripke structure $\M_{int} =
(\mathbb{W}, \R, \D, V)$, where each Montague's possible world $w_n
\in \mathbb{W}$ is the multi-modal translation of a P2P database
system in that world, given by Definition \ref{def:kripke}, that is
%\begin{center}
$~~~w_n = (\W,\{\R_i\},\D,\V)_n~\in \mathbb{W}$,
%\end{center}
so that an intensional equivalence of views, $q_i(\textbf{x})$ and
$q_j(\textbf{x})$, defined as conjunctive queries over peers $P_i$
and $P_j$ respectively, is formally given by the following modal
formulae of the intensional FOL:
\begin{center}
$\lozenge q_i(\textbf{x}) \equiv  \lozenge q_j(\textbf{x})~~$ i.e.,
$~~(\lozenge q_i(\textbf{x}) \Rightarrow  \lozenge q_j(\textbf{x}))
\wedge (\lozenge  q_j(\textbf{x}) \Rightarrow \lozenge
q_i(\textbf{x}))$
\end{center}
\end{definition}
This definition tells us, intuitively, that any possible world (for
a given time-instance) of the intensional logic for P2P database
system $\N$, represents (that is models) a particular state of this
P2P database, that is, the structure and the extensions of all peer
databases in such a time instance. The set of possible worlds $w_n
\in \mathbb{W}$ corresponds to the whole evolution in time of the
given P2P system. Such an evolution is result of all possible
modifications of an initially defined P2P database system:  a simple
modification of extensions of peer databases,  an inserting of a new
peer, or a deleting of an existing peer in this network $\N$.
%Thus, the modal formulae used to specify the intensional equivalence
%of two views, $\langle q_i(\textbf{x})\rangle_{\textbf{x}} \approx
%\langle q_j(\textbf{x})\rangle_{\textbf{x}}$, means that if, for a
%given tuple of constants $\textbf{c}$, $K_i q_i(\textbf{c})$ is true
%in some world $w \in \mathbb{W}$, than also $K_j
%q_j(\textbf{c})$ may be true in some world $w' \in \mathbb{W}$.
%
%
\section{Weak intensional inference  relation}
In real Web applications we will never have  the \emph{omniscient
query agents} that will contemporary have the complete knowledge
about \emph{all} ontologies of all peers. Such a supposition would
generate the system with a global and centralized knowledge, in
contrast with our pragmatic and completely decentralized P2P systems
with completely independent peers, which can change their local
ontology in any instance of time without informing any other peer or
"global" system about it. Thus, what we will consider that a query
agent reasoning system has a weaker form of deduction than $
\vdash_{in}$  (of this ideal omniscient intensional logic
inference), more adequate for  limited and local knowledge of query
agents about the peers. What we consider is that a query agent will
begin its work for a given user query $q(\textbf{x})$ over a peer
$P_i$, and, by using only the \emph{local} knowledge about this
peer's ontology and the set of its local intensional mappings
towards other peers, it will be able to move to the locally-next
peers to obtain answers from them also. This context-sensitive query
answering is analog to the human query answering: interviewer will
ask the indicated person and will obtain his known answer, but this
person can tell also which other people, he believes, will be able
to respond to this question. It will be the task of the interviewer
to find other people and to reformulate the question to them. It is,
practically impossible to have all people who know something about
this question to be in common interaction one with all other to
combine the partial knowledge of each of them in order to provide
possibly complete  answer to such a question. \\
This, context dependent and locally-based query answering system,
for practical query agents in P2P systems, is partially described in
the Example 1. In what follows we will define the \emph{weaker
deductive inference} relation also \cite{Majk09e}, denoted by
$\Vvdash$, for the intensional FOL  $\L_{\omega}$, such that the
query answering algorithm used by these non-omniscient query agents,
is complete
w.r.t. this intensional logic deductive system.\\
%$\vspace*{-3mm}$
\textbf{Example 1:} Let us consider the \emph{cyclic} P2P system in
a Fig.2, with  a sound but generally incomplete deduction
\cite{Majk06Om}, which can be easily implemented by
\emph{non-omniscient} query agents: we have $P_i$, with the ontology
$\O_i$ and the interface $\M^{ij} = \{(v_{im}, v_{jm})~|~ 1\leq m
\leq k_1 \}$ toward the peer $P_j$, and the peer $P_j$, with the
ontology $\O_j$ and the interface $\M^{ji} = \{(w_{jm}, w_{im})~|
~1\leq m \leq n_1 \}$ toward the peer $P_i$. First we traduce  a
pair $(v_{im}, v_{jm})$ by the intensional equivalence $\langle
v_{im}\rangle \approx \langle v_{jm} \rangle$. In what follows, the
subscript of a query identifies the peer relative to such a query.
\begin{figure}
$\vspace*{-21mm}$
\begin{center}
 \includegraphics{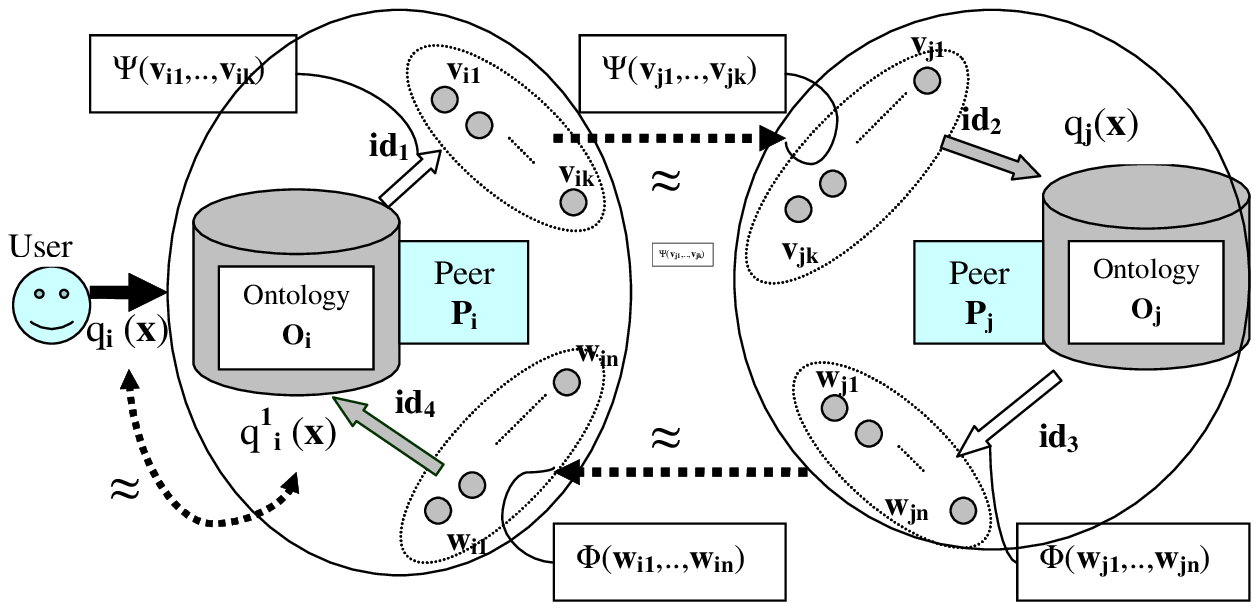}
   \caption{Derivation of intensionally equivalent queries}
  \label{Operations}
 \label{fig:cyclic}
 \end{center}
 $\vspace*{-13mm}$
 \end{figure}
\\Let $q_i(\textbf{x})$ be the original user's conjunctive query over
the ontology $\O_i$ of the peer database $P_i$. If this query can be
rewritten \cite{AlMS95}, by the query rewriting algorithm $id_1$,in
the equal query over the set of views $\{v_{i1},...,v_{ik} \}
\subseteq \pi_1\M^{ij}$, where $\pi_1$ is the first projection, we
will obtain identical (to original query $q_i(\textbf{x})$)
conjunctive query $\Psi(v_{i1},...,v_{ik})$, that is, in the
intensional logic language holds the identity \\$~~id_1: \langle
q_i(\textbf{x}) \rangle_{\textbf{x}} = \langle
\Psi(v_{i1},...,v_{ik}) \rangle_{\textbf{x}}$, or, equivalently,
$~\square(q_i(\textbf{x}) \equiv
\Psi(v_{i1},...,v_{ik}))$.\\
%From the fact that each peer $P_i$ is a S5 normal modal logic with
%the epistemic operator $K_i$, so that holds $K_i(A_1 \wedge ...
%\wedge A_l) = K_i A_1 \wedge ... \wedge K_i A_l$, from the
%conjunctive formula $\Psi(v_{i1},...,v_{ik})$, over views $v_{im}, 1
%\leq m  \leq k$, we obtain that $K_i\Psi(v_{i1},...,v_{ik}) =
%\Psi_1(K_iv_{i1},...,K_iv_{ik})$, and by
From the set of intensional equivalences in $\M^{ij}$, $~~\langle
v_{im} \rangle \approx \langle v_{jm} \rangle, 1 \leq m \leq k$, we
obtain that $~~\langle \Psi(v_{i1},..,v_{ik})\rangle_{\textbf{x}}
\approx \langle \Psi(v_{j1},..,v_{jk})\rangle_{\textbf{x}}$, $~$ or
$~\lozenge \Psi(v_{i1},..,v_{ik})\equiv \lozenge
\Psi(v_{j1},..,v_{jk})$,
%(but the left and right side expressions have not necessarily the same extension in a given possible world),
%\\denoted by $~~~~\Psi(v_{i1},...,v_{ik})\approx\Psi(v_{j1},...,v_{jk})$
in the top-horizontal arrow in Fig.1. \\In the next step the
conjunctive query formula $\Psi(v_{j1},..,v_{jk})$ over the set of
views $\{v_{j1},...,v_{jk} \}\subseteq \pi_1 \M^{ji}$ of the peer
database $P_j$, will be rewritten (by simply unfolding) to the
conjunctive query $q_j(\textbf{x})$ directly over the ontology
$\O_j$ of the peer $P_j$, that is, in the intensional
logic language holds the identity\\
$~~id_2: \langle \Psi(v_{j1},...,v_{jk}) \rangle_{\textbf{x}} =
\langle q_j(\textbf{x}) \rangle_{\textbf{x}}$, or, equivalently,
$~\square(q_j(\textbf{x}) \equiv
\Psi(v_{j1},...,v_{jk}))$.\\
If we compose algebraically these mappings we obtain the one-step
P2P query rewriting $~id_2 \circ \approx \circ id_1 :
q_i(\textbf{x}) \mapsto q_j(\textbf{x})$, that is, from $~id_2 \circ
\approx \circ id_1 ~=~ \approx~$ we obtain the intensional
equivalence $~~~\langle q_j(\textbf{x}) \rangle_{\textbf{x}} \approx
\langle q_i(\textbf{x}) \rangle_{\textbf{x}}$, $~~~$ that is,
$~~\lozenge q_j(\textbf{x})
 \equiv   \lozenge q_i(\textbf{x})$.\\
 In the same way (see the inverse bottom horizontal arrows of a
 diagram in Fig.1), based on the interface specification of the peer $P_j$, $\M^{ji} =
\{(w_{jm}, w_{im})~|~ 1\leq m \leq n \}$, toward the peer $P_i$, we
obtain that also $~~\langle q^1_i(\textbf{x}) \rangle_{\textbf{x}}
\approx \langle q_j(\textbf{x}) \rangle_{\textbf{x}}$, $~~~$ that
is, $~~\lozenge q^1_i(\textbf{x})
 \equiv   \lozenge q_j(\textbf{x})$.
 Thus we obtain the three intensionally equivalent queries
 $q_i(\textbf{x})$, $q_j(\textbf{x})$ and  $q^1_i(\textbf{x})$, where two of them, $q_i(\textbf{x}),
 q^1_i(\textbf{x})$ are over the \emph{same}  peer $P_i$: the first
 one is the original user query, while the second is the
 intensionally equivalent \emph{derived} query (based on P2P
 interface intensional specification).\\
  These three query formulae, $ \{ q_i(\textbf{x}),
 q_j(\textbf{x}),
 q^1_i(\textbf{x})\}$,
 are  the subset of the equivalent class $\C$  for the given user query, which in the
 intensional FOL  $\L_{\omega}/_{\approx}$ is  represented by
 the quotient intensional entity $Q(\textbf{x})$, whose extension  (from
 Definition \ref{def:equival})  in the actual
world $w_0$ is defined by $~~\F(w_o)(den(\langle Q(\textbf{x})
\rangle_{\textbf{x}})) =
\{\textbf{t}\in \D^k~|~Q(\textbf{t}), Q(\textbf{x})\in \C\}~=\\
 = \bigcup_{1\leq i
\leq m} \F(w_0)(den(\langle Q(\textbf{x}) \rangle_{\textbf{x}}))$,
that is,  the union of known answers of  these three queries is a
subset of the extension of this quotient intensional entity
$Q(\textbf{x})$. \\$\square$\\
% Only in the case when this sound  query rewriting
%algorithm deduces \emph{all} intensionally equivalent queries in a
%P2P network, that is when it is \emph{complete} also,  the union of
%known answers for queries obtained from such algorithm will be equal
%to the extension of the intensional entity $Q(\textbf{x})$.
%
%In precedence we discussed the necessity to employ \emph{non
%omniscient} query agents, which can use only some local P2P
%information in order to answer to user queries. In such
%\emph{context-dependent} query answering, the answer to an query
%depends on the topology of the P2P network, that is, on the peer's
%accessibility relation $\R_0$, so that for equivalent queries, but
%formalized over different peers we will generally obtain different
%answers. Thus,
The semantics for this weaker form of deduction of intensional
equivalences, i.e., of the intensional equivalent queries over other
"contextual" peers, w.r.t. the user interrogated peer $P_i$, can be
formally expressed by deduction chains which begin from a peer
$P_i$.
 %$\vspace*{-3mm}$
\begin{propo} \label{def:inference} Given an intensional logic $\L_{\omega}$ for a P2P system (Definition \ref{def:P2Plogic}), with  a basic, user
defined, set of intensional equivalences $S_{eq}$, and its deductive
inference relation $ \vdash_{in}$, then we define the weak
intensional inference relation $\Vvdash$, as follows \\
$~~~~~~~~~~~~~~~~~~~~\L_{\omega}~\Vvdash ~\langle A \rangle_{\alpha}
\approx
\langle B \rangle_{\alpha}~~~~~~~$  iff\\
 there is a  chain $A_1, A_2,A_3...,A_{3n+1}$ of the formulae with
the same set of free variables but each of them expressed by
relation symbols of only one particular peer's ontology, such that
$A_1 = A, ~A_{3n+1} = B$, and $\langle A_i \rangle_{\alpha}
\cong_{i+1} \langle A_{i+1} \rangle_{\alpha}$, for $~\cong_{3i}$
equal to $ \approx$; to $=$ otherwise, $A_{3i-2}$ is a query over a
peer's ontology, and $A_{3i-3}, A_{3i-1}$ over peer's views, while
$A_{3i}$ is
a query over views contained in the interface of this peer but are views of some other peer, $~1 \leq  i  \leq n$.\\
 These chains for the intensional
logic $\L_{\omega}~$ of a P2P database system are finite, and holds
that $~\L_{\omega}~\Vvdash ~\langle A \rangle_{\alpha} \approx
\langle B \rangle_{\alpha}~$ implies $~\L_{\omega}~\vdash_{in}
~\langle A \rangle_{\alpha} \approx \langle B \rangle_{\alpha}~$,
but not vice versa.
%$\vspace*{-4mm}$
\end{propo}
\textbf{Proof:}
%1.From-left-to-right: if $ ~\langle A
%\rangle_{\alpha} =_{in} \langle B \rangle_{\alpha}~$, then this
%chain is just $A,B$. 2. From-right-to-left:
 This proposition tells us that two formulae, over any two peer's
ontologies, with the same free variables, are intensionally
equivalent, if there is a chain of the formulae, identical or
intensionally equivalent, and these two formulae are the initial and
final formulae of such a chain. In fact, any two identical formulae
can be reduced to only one, by eliminating other (substitution
property for identity), that is, we are able to reduce such a chain
to the sub chain with only intensional equivalent formulae, and
based on the transitive property of the equivalence relation, we
obtain that initial and final formula in this chain are
intensionally equivalent.\\
The finite chain property for P2P systems is the result of the fact
that, also in presence of cyclic mappings between peers, the number
of \emph{different} conjunctive queries but intensionally
equivalent, which can be expressed by the finite set $S$ of views of
a peer $P_i$, used as mapping toward the same peer $P_j$ ($i\neq
j$), is always finite. The number of subsets of this set $S$ of
view, sufficient to formalize the intensionally equivalent
conjunctive formula, is finite: thus, the passage from $P_i$ to
$P_j$ during the derivation of new intensional equivalences, can be
used only a finite number of times.
\\$\square$\\
 Moreover, this proposition
explains the way in which weak deduction of the Intensional logic is
able to derive the intensionally equivalent formulae  from the basic
set (explicitly defined by a peer's developers) of intensionally
equivalent formulae: in our case it is the set of intensionally
equivalent views (conjunctive queries) over different peers.
%$\vspace*{-3mm}$
\begin{propo} \label{def:query} Let $~~\L_{\omega}\Vvdash
~\langle  q_i(\textbf{x}) \rangle_{\textbf{x}} \approx \langle
q'_i(\textbf{x}) \rangle_{\textbf{x}}~$ be a weak deduction of the
intensional equivalence, where the bottom index $i$ of the
conjunctive queries denotes the peer $P_i$ relative to these
queries, with its views $v_1,..,v_k$, and, from the Prop.
\ref{def:inference},  $A_1, A_2,...,A_n$ be a finite chain of the
formulae with the same set of free variables, such that $ A_1 \equiv
 q_i(\textbf{x})$, $A_2 \equiv  \Psi(v_1,..,v_k)$, $\langle A_1
\rangle_{\textbf{x}} = \langle A_2\rangle_{\textbf{x}}$ and
$~A_{n-1} \equiv q'_i(\textbf{x})$, $A_{n} \equiv \Phi(v_1,..,v_k)$,
$\langle A_{n-1} \rangle_{\textbf{x}} = \langle
A_{n}\rangle_{\textbf{x}}$, where conjunctive queries $\Psi$ and
$\Phi$ have also the same free variables (view attributes).\\ Then
$~~~~~~~~~\Bbbk(den(\langle \Phi(v_1,..,v_k) \rangle_{\textbf{x}} ))
\subseteq \Bbbk(den(\langle \Psi(v_1,..,v_k) \rangle_{\textbf{x}}
))$.
%\\denoted by an equivalence  $ ~~~~~~~q_i(\textbf{x}) \congq'_i(\textbf{x})$.
% $\vspace*{-5mm}$
\end{propo}
\textbf{Proof:}  It comes directly for all user conjunctive-queries
without built-in predicates: two conjunctive queries with the same
set of predicates and the same set of variables in the query head
are identical. In the case when user query contains also a derived
built-in predicate $C(\textbf{x})$, we can take out this formula
from the rest of query, and consider the intensional equivalence
only for such reduct without built-in predicates, based on the
Proposition  \ref{prop:equival}: at the end of derivation of  the
intensional equivalence class w.r.t. this reduct query, we can add
to each conjunctive formula of this equivalence class the formula "$
~\wedge~ C(\textbf{x})$". The chain of derivations can only add some
new conjunction of built-in predicates, thus we will obtain that
$\Psi(\textbf{x}) \equiv \Phi(\textbf{x}) \wedge C_1(\textbf{x})$,
 where $C_1(\textbf{x})$ can be also empty. \\$\square$\\
 This proposition tells us that any two intensionally
equivalent conjunctive queries, with the same set of virtual
predicates (views $v_1,..,v_k$ of a peer $P_i$) and the same set of
variables in the head of these two queries, the second derived query
is extensionally contained in the first, so that we can stop the
propagation of deductions and to discard $\Psi(\textbf{x})$.
 As a consequence of Propositions  \ref{def:inference} and
 \ref{def:query},
given a query $q(\textbf{x})$ over a peer $P_i$, the set of
\emph{different} conjunctive queries (such that one is not subsumed
in other), but intensionally equivalent to $q(\textbf{x})$, over any
peer $P_k$ is a finite set: that means that in principle we are able
to define a complete query rewriting algorithm for finite $P2P$
database systems w.r.t. the weak deduction $\Vvdash$ of the
intensional FOL $~\L_{\omega}$. More about this non omniscient
inference can be found in \cite{Majk06Om} also.
%\label{prop:equival}
 %
\section{Sound and complete query answering}
The implementation of query answering in P2P systems  needs a
standard mathematical semantics based on an adequate (co)algebra: as
for example,is the \emph{relational} (co)algebra for SQL query
answering in Relational Databases. Here, the computation is more
intricate because of the complex epistemic logic structures of peers
and the necessity of query rewriting algorithms $Rew$. We consider
that the rule of query agent is to start and to maintain complete
\emph{query answering transaction}: this transaction starts when is
defined an user query $q(\textbf{x})$ over a peer $P_i$. A query
agent supports the $Rew$ algorithm in order to construct
intensionally equivalent rewritten queries over other peers and then
calls grid computation network to calculate answers, by assigning to
each grid computation node one peer with a \emph{union} of rewritten
conjunctive queries for it. The transaction ends when query agent
receives the answers from all grid nodes, and presents collected
answers to the user. The definition of this P2P query computing
system, which abstracts all not necessary implementation details of
a peer, has to be given in an abstract (co)algebraic mathematical
language; so, this abstract mathematical specification (co-Algebraic
Abstract Type) can be successively
implemented in any current grid computing system.\\
But the query answering for intensionally based P2P mappings can not
be embedded into recursive Datalog, as in the case of a standard
view-based mappings based on a material implication \cite{CDGL04},
so we need more complex and general mathematical framework for it.
%$\vspace*{-3mm}$
%
\subsection{Final semantics for the weak deduction $~~\Vvdash$}
\begin{figure}
$\vspace*{-7mm}$
\begin{center}
 \includegraphics{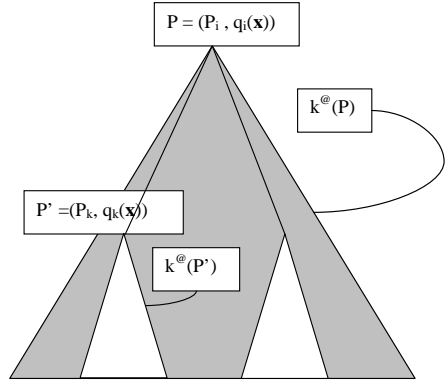}
   \caption{Weak deduction solution}
   \label{fig:treeinf}
  \end{center}
  $\vspace*{-11mm}$
 \end{figure}

The Kripke structure of the frame $\F = (\W,\{\R_i\})$, given in the
Definition \ref{def:kripke}, is a prerequisite in order to obtain a
\emph{coalgebraic semantics} for query answering in P2P database
framework.  (Co)Algebras provide an unifying view on a large variety
of dynamic systems such as transition systems, automata, data
structures, and objects ~\cite{Jaco96,Rute00} or Kripke models; they
are especially useful for the dynamic query answering P2P systems.
In order to render this paper more selfcontained we will introduce
the following formal definition for coalgebras:
%$\vspace*{-3mm}$
\begin{definition} (Abstract Coalgebras) Let $Set$ be a category with its
objects all (small) sets and its arrows all functions, and $T$ be an
endofunctor (mapping) $T:Set\rightarrow Set$.  A T-coalgebra is a
pair of a (small) set $C$ and a Set-arrow $\alpha:C \rightarrow TC$,
that is, $ (C,\alpha:C \rightarrow TC)$. $~~~T~$ is called a
signature functor or type, and $C$ a carrier set. Let $(C,\alpha)$
and $(D,\beta)$ be T-coalgebras, and $f:C \rightarrow D$ a
Set-arrow. $f$ is said to be a morphism of T-coalgebras or
T-morphism, if $~~\beta \circ f = Tf \circ \alpha$, where $\circ$ is
a composition of arrows. It is an isomorphism if it is a bijective
mapping.
%$\vspace*{-8mm}$
\end{definition}
\textbf{Example 2}: Aczel's semantics of CCS \cite{Acze88}, is
described by the coalgebra $k:Prog \rightarrow \P_{fin}(Act \times
Prog)$, of the endofunctor $ T = \P_{fin}(Act \times \_)$ with the
set of actions
 $a \in Act$, such that $k(P) = \{<a,P'>~|~
P \rightarrow_{a_i} P'\}$ is the set of atomic transitions which the
CCS program P can perform and pass to the new program P'. The symbol
$\P_{fin}$ is the finite powerset operator. This semantics exploits
the special final coalgebra theorem, that is a unique homomorphism
$k^@:(Prog, k) \rightarrow (gfp(T), \simeq)$ to the \emph{final
coalgebra}, which is a isomorphic (bijective) coalgebra:
$\simeq:gfp(T)\rightarrow \P_{fin}(Act \times Prog)$ with $gfp(T)$
the set of (infinite) labeled transition systems (labeled trees)
which are greatest fixed points of the 'behavioral functor' $ T =
\P_{fin}(Act \times \_)$, that is for every program P, $~~k^@(P) =
\{<a,k^@(P')>~|~ P \rightarrow_{a_i} P'\}$, such that the following
diagram commutes
%$\vspace*{-2mm}$
\begin{diagram}
   Prog  & \rTo^{k^@}   & gfp(T) \\
 \dTo^{k} &       &   \dTo_{\simeq}\\
 T(Prog)  &  \rTo^{T(k^@)}    &    T(gfp(T))
 %$\vspace*{-1mm}$
 \end{diagram}
 Final coalgebras are 'strongly extensional', that is, two elements
 of the final coalgebra are equal iff they are $T-bisimilar$.
 \\$\square$\\
 The semantics above for CCS and its properties can be generalized
 to arbitrary behaviors: in our case, for a given
conjunctive query language $\L_Q$ over relational symbols of P2P
database system, we consider the programs as pairs $(P_i,
q_i(\textbf{x}))$ (a query over a peer $P_i$ can be considered as
 a program: the execution of this program will return the known answers to this query) so that $Prog = \W_0 \times \L_Q$,
 the set $Act = \{\Vvdash_3\}$ as a singleton,
  with the only deductive action $\Vvdash_3$ ( restriction of $\Vvdash$ for chains of length 3 only),  so that $\P_{fin}(Act \times Prog)$ can be substituted by $\P_{fin}(Act \times Prog)$, that is,
  in our case $T = \P_{fin}$.\\
 Thus, we can consider the \emph{intensional deduction} process, defined in Proposition \ref{def:inference},as a
coalgebra $k:\W_0 \times \L_Q \rightarrow \P_{fin}(\W_0 \times
\L_Q)$, such that, given an initial query $q_i(\textbf{x}) =
q(\textbf{x}) \in \L_Q$ over a peer $P_i$, that is a pair
$(P_i,q(\textbf{x})) \in \W_0 \times \L_Q$, which corresponds to the
intensional entity $\langle q(\textbf{x}) \rangle_{\textbf{x}}$, the
inferential step will generate the complete set
$k(P_i,q(\textbf{x})) = \{(P_j, q_j(\textbf{x}))~|~(P_i,P_j) \in
\R_0  \}~~ \in \P_{fin}(\W_0 \times \L_Q)$, where $\langle
q(\textbf{x}) \rangle_{\textbf{x}} \approx \langle q_j(\textbf{x})
\rangle_{\textbf{x}}$ if it can be derived as an
intensional-equivalent query $q_j(\textbf{x})$ over a peer $P_j$
(that is if $~~\L_{\omega}~\Vvdash_3 ~\langle q(\textbf{x})
\rangle_{\textbf{x}} \approx \langle q_j(\textbf{x})
\rangle_{\textbf{x}}~$) ; $q_j(\textbf{x}) = \emptyset$,
otherwise.\\
We can use $\P_{fin}$ because P2P system is composed by a finite
number $N$ of peers, so that for any peer $P_i$ the number of
accessible peers for it is finite.\\ By  applying recursively the
function $k$, equivalent to the single application of the
\emph{unique} homomorphism $k^@$, we obtain a possibly infinite
transition relation (tree), see Fig. \ref{fig:treeinf}, with the
root in the initial state (that is, a program $(P_i,q(\textbf{x})$).
We consider the general case of cyclic mappings of the P2P system as
in Fig. \ref{fig:cyclic} of the example 1.
\\
This tree will have a lot of nodes with empty queries, and possibly
infinite copies of nodes (with the same query over a given peer).
So, we need to 'normalize' this unique solution of the weak
deduction, by eliminating duplicates and nodes with empty query.\\
Consequently we define the mapping $fl:gfp(\P_{fin}) \rightarrow
\P_{fin}(\W_0 \times \P(\L_Q))$, such that, given a unique solution
(tree)
$k^@(P_i, q_i^1(\textbf{x}))$ , \\
$fl(k^@(P_i, q_i(\textbf{x}))) =
% \{(P_k, \{q_k^1(\textbf{x}),...,
%q_k^n(\textbf{x}),...\})~|~(P_k, q_k^m(\textbf{x})) \in k^@(P_i,
%q_i(\textbf{x})),\\ 1 \leq k \leq N, m \in \mathbb{N} \}$,
 \bigcup_{1 \leq k \leq N}(P_k, \{q_k^m(\textbf{x})~|~(P_k, q_k^m(\textbf{x})) \in k^@(P_i,
q_i(\textbf{x})), m \in \mathbb{I} \})$,
 where
$\mathbb{I}$ is the set of integers. In this way we will obtain, for
each peer, the set of all
 conjunctive queries, intensionally equivalent to the user query
 $q_i^1(\textbf{x}) = q(\textbf{x})$. This set of queries for a given peer is complete
  w.r.t. the intensional FOL  and its weak deductive
 inference for intensionally equivalent formulae. It is the largest equivalence relation class, i.e.,
 the closure $C_{\Vvdash}(\L_{\omega}, q_i^1(\textbf{x}))$ of the inference $\Vvdash$,
 defined by $q'(\textbf{x}) \in C_{\Vvdash}(\L_{\omega}, q_i^1(\textbf{x}))~$ iff $~\L_{\omega} \Vvdash \langle q_i^1(\textbf{x})
\rangle_{\textbf{x}} \approx \langle q'(\textbf{x})
\rangle_{\textbf{x}}$.
 %$\vspace*{-3mm}$
 %
\subsection{Sound and complete w.r.t $~\Vvdash ~$ query answering algorithm}
 In order to have a decidable complete query answering,
 we need to prove that the set of all rewritten queries for each peer is
 finite. After that we need to define this query answering
 algorithm $Rew$ and to prove that it terminates, and returns with the
 same solution as the unique solution of  the
 final coalgebra for the weak deduction.\\
 In what follows we will define the query agent in an
 object-oriented style, as a class, described as coalgebra
 \cite{Jaco96} $st:X \rightarrow B+X$ of a deterministic system for a polynomial functor
  $T = B+ \_~~$ where $B = \P_{fin}(\W_0\times \P_{fin}(\L_Q))$ is the set of
termination values of query rewriting algorithm, defined as the set
of finite queries for each peer.  $X = (Ls_Q \times
\mathbb{N})^{\W_0}$ is the set of internal states of this query
agent class: each state is the
 function $f:\W_0 \rightarrow Ls_Q \times \mathbb{N}$, which maps any peer $P_i \in \W_0$
 into a list of conjunctive queries over this peer in $Ls_Q$ ( $Ls_Q$ is the list
 of all possible conjunctive queries over the alphabet of the P2P system),
 and to the pointer $k \geq 0$ in $\mathbb{N}$ which defines the position of the last
 elaborated query for this peer. Such a query agent class is instantiated into a query agent \emph{object} using "new" operation,
  performed by an user when he defines the query $q(\textbf{x})$ over a peer $P_i$, denoted by the mapping $new:\textbf{1} \rightarrow X$,
  where $\textbf{1} = \{ * \}$ is the singleton set. The operation "new" specifies the
 initial state $f_0$ of the object, $f_0 = new(*) \in X$: it is a function $f_0:\W_0 \rightarrow Ls_Q \times \mathbb{N}$,
  such that for a given user query $q_i(\textbf{x})$ over a peer
 $P_i$, $f_0(P_k) = (<q^1_i(\textbf{x})>,0)$, if $k = i$; $~~(<\emptyset>,0) $ otherwise,
 where $\emptyset$ denotes empty query.\\ With $ln$ we denote the function which
 returns with a number of elements in a list S, while with $push(S,q)$ the function which inserts the element $q$ as the last element of this list.
  If $\pi_m, ~m= 1,2,..$ denotes the m-th projection,  then for any peer $P_k$, $\pi_1f(P_k)$
 is the list of queries for this peer, and  $\pi_2f(P_k)$ is the pointer to the last elaborated query in this list.
  Thus,  every query agent object
 (instance of this object-oriented class) can be denoted as a couple $(st, f_0)$.\\
 The polynomial functor $T(X)
= B+X$, where $+$ is the operation of the set union,  has the final
coalgebra semantics \cite{Jaco96}, with $gfp(B+ \_~) =
\P_{fin}(\W_0\times \P(\L_Q))$, the infinite extension of $B$ when
for each peer in $W_0$ we can have also infinite number of queries,
so that $ B+ gfp(B+ \_~) = gfp(B+ \_~)$ and holds the bijection
$\simeq$ for the final coalgebra of the functor $T = B+ \_~~$. We
denote the unique solution, of the deterministic system $st:X
\rightarrow B+ X$, by the unique homomorphism between this coalgebra
and the final coalgebra,
that is, $st^@:(X,st) \rightarrow (gfp(B+ \_ ~),~ \simeq)$; this homomorphism corresponds to the top commutative diagram in the diagram below.\\
 The coalgebra mapping, $st$, specifies the \emph{method}
 of this query agent class for the query rewriting \emph{algorithm}, as
 follows: given an initial state $f_0 \in X$,  the $st$ terminates with a result of query rewriting algorithm
 if all queries of every peer are elaborated, and are not generated new queries;
 % we will see  that $st$ is a monotonic operator so that it terminates with a fixpoint solution $st^@(f_0)$;
 otherwise $st$, when elaborates a first non-elaborated query $q_k^m(\textbf{x})$
 of a peer $P_k$, can generate the set of \emph{new} intensionally equivalent queries over other peers,
  and passes (deterministically) to the next state in $X =(Ls_Q \times \mathbb{N})^{\W_0} $. \\
  Formally, for any state $f:\W_0 \rightarrow Ls_Q \times
  \mathbb{N}$,\\
  $\vspace*{-2mm}$
  $----------------------------------------$\\
 %\hline
 %$\vspace*{-2mm}$ \\\\
  $st(f) = $ \\
   $= \{(P_k, \{q_k^n ~|~q_k^n \in \pi_1f(P_k), 1 \leq n \leq \pi_2f(P_k)\} ) ~|~ P_k \in
  \W_0\}$\\
   $~~~~~~$ if $~~~~~~0 = \sum_{P_i \in \W_0} ln(\pi_1f(P_i)) -
   \pi_2f(P_i)$,\\
$= next(f) = f_1:\W_0 \rightarrow Ls_Q \times \mathbb{N} ~~~~~~$
otherwise, $~~~~~~~~~$
such that\\
$~~~f_1(P_k) = (S,n+1)$, with $(S,n) = f(P_k)$ : increments
pointer.\\
$~~~$ For any  $P_j$ locally connected with  $P_k$, and $(S,m) = f(P_j)$,\\
$~~~f_1(P_j) = (push(S,q_j^{l+1}),m)$, for $ l =
ln(S)$\\
 $~~~~~~~~~~~~~~~~~~$ if $ q_j^{l+1} =Rew( \pi_n(\pi_1f(P_k)), P_j) \neq
 \emptyset$ $~~~~~~$
 and\\
 $~~~~~~~~~~~~~~~~~~~~~~~~~~~~\forall_{1\leq i \leq l}~\neg(q_j^{l+1}\cong \pi_i(\pi_1f(P_j)))~~~~~$ (Proposition \ref{def:query})\\
$~~~~~~~~~~~~~~~~~~$ otherwise $f_1(P_j) =
f(P_j)$.\\
%$\vspace*{-3mm}$
%\hline
 $----------------------------------------$
 $\vspace*{-2mm}$\\\\
  The peer-to-peer query step-rewriting $Rew = Unfolding \circ Subst \circ Minicon$,
  where $\circ$ is the sequential composition for algorithms, can be described as follows (see also Example 1):\\
  Given a conjunctive query
  $q(\textbf{x})$ over a peer $P_i$, by using the set of intensional
  equivalences in its interface $\M^{ij} = \{(v_{im}, v_{jm})~|
~~1\leq m \leq k_1 \}$ toward a peer $P_j$ (see Def.
\ref{def:P2Pnetwork}), in the case when such set is not enough for
the complete and \emph{equivalent} rewriting \cite{Hale01}, returns
with the empty query $q_j(\textbf{x}) = \emptyset$ for the peer
$P_j$. Otherwise it uses the MiniCon Algorithm \cite{Hale01} over
the set of views in $\{v_{i1},...,v_{ik} \} \subseteq \pi_1\M^{ij}$,
to  rewrite equivalently a query $q(\textbf{x})$ into a query
$\Psi(v_{i1},...,v_{ik})$. After that it makes the
 substitution of views of $P_i$ in
$\{v_{i1},...,v_{ik} \}$  by intensionally equivalent set of views
of $P_j$ in $\{v_{i1},...,v_{ik} \}$, to obtain an intensionally
equivalent query formula $\Phi(v_{i1},...,v_{ik})$ over views of
$P_j$. Finally, it uses the Unfolding Algorithm \cite{Hale01}, to
unfold $\Phi(v_{i1},...,v_{ik})$ and to obtain the query
$q_j(\textbf{x})$ over the ontology of a peer $P_j$. Notice that in
the case when the ontology of $P_j$ is changed, so that the set of
views $v_{i1},...,v_{ik}$ in the interface $\M^{ij}$ of the peer
$P_i$ does not match with this new ontology of $P_j$, the algorithm
returns with the empty query, that is, with $q_j(\textbf{x}) =
\emptyset$.
%$\vspace*{-3mm}$
\begin{propo} The mapping $next:X \rightarrow X$ is monotonic w.r.t. the ordering $\preceq$ such
that for any $~~f_1,f_2 \in X = (Ls_Q \times \mathbb{N})^{\W_0}$,\\
$~~~~~~~~~f_1 \preceq f_2~~~$ iff $~~~\forall P_i \in \W_0
(\pi_1(f_1(P_i)) \subseteq
\pi_1(f_2(P_i)))$.\\
For the least fixpint for this "next-consequence-operator" of
 $f_0 = new(*)$, $~~lst(f_0)$, holds that
$~~st^@(f_0) = st(lst(f_0))$.
% $\vspace*{-4mm}$
\end{propo}
 Notice that this algorithm works well
also for \emph{union} of conjunctive queries: it works well for
$Unfolding$ and $MiniCon$ \cite{Hale01}, while for $Subst$ works
from the fact that for a normal modal logic holds $\lozenge(A \vee
B) \equiv  \lozenge A \vee \lozenge B$. The $Subst$ works for
conjunctive queries and the class of peers defined as follows:
%$\vspace*{-3mm}$
\begin{propo} \cite{Majk04ph} Let us consider the class of peers with the integrity
constraints which does not contain negative clauses of the form
$\neg A_1 \vee ...\vee \neg A_m, ~m \geq 2$. Then, the intensional
equivalence is preserved by conjunction logic operation, that
is,\\
if $~~~\varphi \equiv (b_1 \wedge ...\wedge b_k)$, $k\geq 1$, is a
conjunctive query over a peer $P_i$,  and $b_i \approx c_i$, $~1\leq
i \leq k$, the set of intensionally equivalent views toward a peer
$P_j$, $~~~$ then $~~~\varphi\approx \psi$ $~~~$, where $~\equiv~$
is a logic equivalence and $~~~\psi \equiv (c_1 \wedge ....\wedge
c_k)$ is the conjunctive query over a peer $P_j$.
% $\vspace*{-4mm}$
\end{propo}
We are able to define the mapping $pop:X \rightarrow \W_0\times \L_Q
$ between domains of the query agent class coalgebra and deductive
coalgebra, such that, for any $f \in X$, that is, $f:\W_0
\rightarrow Ls_Q \times \mathbb{N}$, $~~~~pop(f) = \{(P_i,
q_i)~|~\P_i
\in \W_0$, and $ q_i =\pi_n(\pi_1(f(P_i)))~~$ if\\
$~~~~~~~~~~~~~~~~~~n = 1+ \pi_2(f(P_i))
\leq ln(\pi_1(f(P_i));~~\emptyset ~~$ otherwise $\}$.\\
This mapping associate to any peer $P_i$ the next query in its list
to be elaborated.
% $\vspace*{-3mm}$
%
\begin{theo} The query answering algorithm implemented by the query
class $st:X \rightarrow B+X$ will terminate for any user conjunctive
query $q(\textbf{x})$ over a peer $P_i$. It is sound and complete
algorithm w.r.t. the weak deduction inference $\Vvdash$ of the
intensional logic $\L_{\omega}$ for a P2P database system.
That is, for any user query action $new$, holds that $~~~~~~~~~~~~~~~~~~~~~~st^@ \circ new = fl\circ k^@ \circ pop \circ new$\\
%$~~~~~~~~~~~~~~~~~~st^@ \circ new = fl\circ k^@ \circ pop \circ new$,\\
or, equivalently, any user action $new$, which defines a conjunctive
query $q(\textbf{x})$ over a peer $P_i$, is \textsc{equalizer} of
the functions $st^@$ and $fl\circ k^@ \circ pop$. Graphically
%$\vspace*{-3mm}$
 \begin{diagram}
 \textbf{1}& \rTo^{new} & (Ls_Q \times \mathbb{N})^{\W_0} &  \pile{ \rTo^{st^@} \\ \rTo_{fl\circ k^@ \circ pop} }  &  ~~gfp(B+ \_~)
 \end{diagram}
% $\vspace*{-7mm}$
\end{theo}
This theorem can be represented by the following commutative
diagram: the top commutative square corresponds to the query agent
with the (unique) solution for its query rewriting algorithm, while
the bottom commutative square corresponds to the unique solution of
the weak deduction inference. The dashed diagram in the middle
corresponds to the equalizer of this theorem, and, intuitively,
shows that each unique solution of query answering algorithm is
equal to the unique solution set obtained by the weak deductive
inference $\Vvdash$ of the intensional logic $\L_{\omega}$ for a P2P
database system.
%\newarrow{Dashto} {}{dash}{}{dash}{\rTo}
\newarrow{Dashto} ....>
 \begin{diagram}
 & & B+(Ls_Q \times \mathbb{N})^{\W_0} &   \rTo^{1_B+st^@} &  B+ gfp(B+ \_~)\\
 & & \uTo^{st} &         &  \uTo_{\simeq} \\
 \textbf{1} &\rDashto^{new} & (Ls_Q \times \mathbb{N})^{\W_0} &   \rTo^{st^@} &  gfp(B+ \_~)\\
& & \dDashto^{pop}   &  &   \uDashto^{fl} &   \\
&  &  \W_0\times \L_Q  & \rTo^{k^@} &   gfp(\P_{fin})& \\
 & &\dTo^{k} &       &   \dTo_{\simeq} &\\
&  &\P_{fin}(\W_0\times \L_Q)  &  \rTo^{\P_{fin}(k^@)}    &
\P_{fin}(gfp(\P_{fin})) &
 \end{diagram}
\textbf{Proof:} The query rewriting algorithm is sound, because it
derives intensionally equivalent queries. From the fact that it
derives the subset of the intensionally equivalent queries over
peers, w.r.t. the weak deductive inference $\Vvdash$ which for a
given finite P2P system derives only a \emph{finite} number of
equivalent queries (from Propositions \ref{def:inference},
\ref{def:query}), we conclude that it must terminate. Let us sketch
now the completeness proof for a given user action $new$, which
specifies a query $q_i$ over a peer $P_i$, such that $f_0 = new(*)$,
and $(P_i,q_i) = pop(f_0)$. We can focus only on non empty queries,\\
1.  Let $(P_k,q_k), q_k \neq \emptyset$ be a node in the infinite
tree  $k@(P_i,q_i) \in gfp(\P_{fin})$. So, there is a chain (path)
from the root of this tree $(P_i,q_i)$ to this node: the set of
mutually different nodes with non-empty queries in this path must be
finite number $n$. Thus, there is a maximal number $m \leq n$ of
consecutive executions of the query rewriting method $st$, denoted
by $st^m$, so that $q_k \in \pi_1(f(P_k))$ for $f = st^m(f_0)$, so,
there exists $(P_k,S) \in st^@(f_0)$ (a unique solution for the user
query in $f_0$, with
$f_0(P_i) = q_i$) such that $q_k \in S$.\\
2. Vice versa, let $(P_k,S) \in st^@(f_0)$ (a unique solution for
the user query in $f_0$, with $f_0(P_i) = \{ q_i \}$) such that $q_k
\in S$. Let prove that $(P_k,q_k)$ must be a node in the  tree
$k@(P_i,q_i) \in
gfp(\P_{fin})$:\\
From the fact that $(P_k,S) \in st^@(f_0)$ we conclude that there
exists a finite number $n$ such that $f_n = st^n(f_0)$ and $q_k =
\pi_m( \pi_1(f(P_k)))$, with $m = ln(\pi_1(f(P_k)))$. Thus there
exists the following sequence-ordered subset of all recursive
executions of the query algorithm method $st$ which begins from
$f_0$ and ends with $f_n$, inductively defined in the backward
direction: the step, immediately precedent to the step $n$ in this
subset, is a step $m_1 \leq n-1$ in which the method $st$ invokes
the action $push(S,q_k)$ for a locally-connected peer $P_j$, i.e.,
$f_{m_1}(P_j) = (push(S,q_k), m')$, which inserts the query $q_k$ in
the list $S$ of the peer $P_k$. Thus, also for some step $m_2 \leq
m_1-1$  the method $st$ invokes the action $push(S',q_j)$ for a
locally-connected peer $P_l$ to $P_j$, i.e., $f_{m_2}(P_l) =
(push(S',q_j), m")$, which inserts the query $q_j$ in the list $S'$
of the peer $P_j$, etc.. In this kind of a backward recursion we
will reach the
beginning element $f_0$ for the initial query couple $(P_i,q_i)$.\\
The chain of nodes $C = <(P_i,q_i), ..., (P_j, q_j), (P_k,q_k)>$ is
a chain of intensionally equivalent queries, thus, must be a weak
deduction inference chain, and, consequently, a part of the unique
derivation tree $~k^@(P_i,q_i)~$ (with the root in the node
$(P_i,q_i)$)). Consequently,  $(P_k,q_k)$ is a node in this tree.
% $\vspace*{-2mm}$
%----------------------------------------------------------------------
%
%----------------------------------------------------------------------
% SECTION VII: Conclusions
%----------------------------------------------------------------------
\section{Conclusion}
 As this paper has shown, the problem of defining the
semantics for intensional ontology  mappings between peer databases,
can be expressed in an intensional FOL language. The
extensionalization function, for the actual P2P world, can be
computationally modeled by an epistemic logic of each peer database:
P2P answers to conjunctive queries are based on the known answers of
peers to intensionally equivalent queries over them. This
intensional FOL for P2P system is obtained by the particular fusion
of the Bealer's intensional algebra and Montague's possible-worlds
modal logic for the semantics of the natural language. In this paper
we enriched such a logic framework by a kind of intensional
equivalence, which can be used to define an intensional view-based
mapping between peer's local ontologies. We conclude that the
intensional FOL logic is a good candidate language for specification
of such P2P database systems. \\We have shown how such intensional
mapping can be used during a query answering process, but we do not
use the omniscient inference of this modal S5 Montague's intensional
logic. Such inference would use all possible worlds, thus would be
very hard to obtain. Rather than it, we defined a \emph{weak}
non-omniscient inference, presented in \cite{Majk06Om} also, which
can be computed in the \emph{actual} Montague's world only, and
based on query-rewriting algorithms: Minicon and unfolding
algorithms for conjunctive queries.\\
% and we define the global existential modal operator $\K$ for a query answering in a
%P2P database system. The semantics of this global modal P2P system
%operator $\K$ is modeled by a kind of the accessibility relation
%between peers, based on the intensional view-based mappings between
%peers, and on the particular sound query rewriting algorithm.\\
% We defined the weak
%deduction inference for it, presented in \cite{Majk06Om} also,
% which
Finaly we show how this query-rewriting algorithm can be
conveniently implemented by non omniscient query agents, and we have
shown that for any P2P system it has a unique final semantics
solution.
%Each particular
%peer database $P_i$ can be seen as a local epistemic S5 modal logic
%with a proper epistemic modal operator $K_i$, "Peer $P_i$ knows that
%..", independently from all other peers, and can be transated as a
%module for a grid computation node during query answering transaction.
It is the responsibility of a query agent to rewrite the original
user query over an initial peer  to all other intensionally
equivalent queries over other peers in a P2P network. We defined the
object-oriented class for such query agents and we have shown that
its query rewriting method (algorithm) is sound and complete w.r.t.
the weak deduction of the intensional FOL.

%----------------------------------------------------------------------

%\bibliographystyle{abbrv}
\bibliographystyle{IEEEbib}
\bibliography{medium-string,krdb,mydb}

%\newpage
%\input{hybridtheory}
%\balancecolumns

%$\vspace*{-4mm}$
\end{document}

%% file: macros-general1.tex
\begingroup
\catcode`\~=11
\gdef\urltilde{\lower 0.6ex\hbox{~}}
\endgroup

\newcommand{\A}{\mathcal{A}} 
\newcommand{\C}{\mathcal{C}} \newcommand{\D}{\mathcal{D}}
\newcommand{\E}{\mathcal{E}} \newcommand{\F}{\mathcal{F}}
\newcommand{\G}{\mathcal{G}} 
\newcommand{\I}{\mathcal{I}} 
\newcommand{\K}{\mathcal{K}} \renewcommand{\L}{\mathcal{L}}
\newcommand{\M}{\mathcal{M}} \newcommand{\N}{\mathcal{N}}
\renewcommand{\O}{\mathcal{O}} \renewcommand{\P}{\mathcal{P}}
 \newcommand{\R}{\mathcal{R}}
\renewcommand{\S}{\mathcal{S}} 
 \newcommand{\V}{\mathcal{V}}
\newcommand{\W}{\mathcal{W}}